\documentclass[11pt]{wlscirep}
\usepackage{times}
\usepackage{graphicx,psfrag}
\usepackage{amsmath,lipsum}
\usepackage{hyperref}
\usepackage[font=sf]{caption}
\usepackage{pdfpages}
\usepackage{threeparttable}
\usepackage{longtable}
\usepackage{CJK}
\usepackage[whole]{bxcjkjatype} 
\usepackage{CJKutf8}
\def\apj{Astrophys.~J.}
\def\apjl{Astrophys.~J.}
\def\apjs{Astrophys.~J.~Suppl.}
\def\mnras{Mon.~Not.~R.~Astron.~Soc.}

\def\nat{Nature}

\def\aj{Astron.~J.}
\def\aap{Astron.~Astrophys.}

\makeatletter
\renewcommand{\maketitle}{\bgroup\setlength{\parindent}{0pt}
\begin{flushleft}
  \textbf{\LARGE \@title}
  \\
  \vspace{0.5cm}
  \@author
\end{flushleft}\egroup
}
\makeatother

\title{A shock flash breaking out of a dusty red supergiant
} 
\date{}

\author
{\large{Gaici Li$^{1,\dagger}$, 
Maokai Hu$^{2,\dagger}$, 
Wenxiong Li$^{3,4, \dagger}$, 
Yi Yang\begin{CJK}{UTF8}{gbsn}
(杨轶)
\end{CJK}$^{5, \ddagger, \dagger}$, 
Xiaofeng Wang$^{1,6,2*}$,
Shengyu Yan$^{1}$,
Lei Hu$^{2,7}$,
Jujia Zhang$^{8,9,10}$,
Yiming Mao$^{11}$,
Henrik Riise$^{12}$,
Xing Gao$^{13}$,
Tianrui Sun$^{2}$,
Jialian Liu$^{1}$,
Dingrong Xiong$^{8,9}$,
Lifan Wang$^{14}$,
Jun Mo$^{1}$,
Abdusamatjan Iskandar$^{13,15}$,
Gaobo Xi$^{1}$,
Danfeng Xiang$^{1}$,
Lingzhi Wang$^{16,4}$,
Guoyou Sun$^{17}$,
Keming Zhang$^{5}$,
Jian Chen$^{2}$,
Weili Lin$^{1}$,
Fangzhou Guo$^{1}$,
Qichun Liu$^{1}$,
Guangyao Cai$^{17}$,
Wenjie Zhou$^{17}$,
Jingyuan Zhao$^{17}$,
Jin Chen$^{17}$,
Xin Zheng$^{17}$,
Keying Li$^{17}$,
Mi Zhang$^{17}$,
Shijun Xu$^{17}$,
Xiaodong Lyu$^{17}$,
A. J. Castro-Tirado$^{18,19}$,
Vasilii Chufarin$^{20,21}$,
Nikolay Potapov$^{22}$,
Ivan Ionov$^{23}$,
Stanislav Korotkiy$^{22}$,
Sergey Nazarov$^{24}$,
Kirill Sokolovsky$^{25,26}$,
Norman Hamann$^{27}$,
Eliot Herman$^{28}$
}
\\
  \vspace{0.5cm}
\normalsize
$^{1}${Physics Department, Tsinghua University, Beijing, 100084, China}\\
$^{2}${Purple Mountain Observatory, Chinese Academy of Sciences(CAS), Nanjing, 210023, China}\\
$^{3}${The School of Physics and Astronomy, Tel Aviv University, Tel Aviv, 69978, Israel}\\
$^{4}${Key Laboratory of Optical Astronomy, National Astronomical Observatories of China, CAS, Beijing 100101, China}\\
$^{5}${Department of Astronomy, University of California, Berkeley, CA 94720-3411, USA}\\
$^{6}${Beijing Planetarium, Beijing Academy of Science and Technology, Beijing, 100044, China}\\
$^{7}${McWilliams Center for Cosmology, Department of Physics, Carnegie Mellon University, 5000 Forbes Ave, Pittsburgh, 15213, PA, USA}\\
$^{8}${Yunnan Observatories, CAS, Kunming, 650216, China}\\
$^{9}${Key Laboratory for the Structure and Evolution of Celestial Objects, CAS, Kunming, 650216, China}\\
$^{10}${International Centre of Supernovae, Yunnan Key Laboratory, Kunming, 650216, China}\\
$^{11}${National Astronomical Observatories of China, CAS, Beijing, 100101, China}\\
$^{12}${Skjeivik Observatory, Strand, 4110 FORSAND, Norway}\\
$^{13}${Xinjiang Astronomical Observatory, CAS, Urumqi, 830011, China}\\
$^{14}${Mitchell Institute for Fundamental Physics and Astronomy, Texas A$\&$M University, College Station, TX 77843, USA}\\
$^{15}${School of Astronomy and Space Science, University of CAS, Beijing, 100049, China}\\
$^{16}${South America Center for Astronomy, National Astronomical Observatories, CAS, Beijing, 100101, China}\\
$^{17}${Xingming Observatory, Urumqi, 830000, China}\\
$^{18}${Instituto de Astrofisica de Andalucia (IAA-CSIC), Glorieta de la Astronomia s/n, Granada, E-18008, Spain}\\
$^{19}${Unidad Asociada al CSIC, Departamento de Ingenieria de Sistemas y Automatica, Escuela de Ingenierias, Universidad de Malaga, Malaga, 29071, Spain}\\
$^{20}${G. M. Grechko Nizhny Novgorod Planetarium, Revolutsionnaya st. 20, Nizhny Novgorod 603002, Russia}\\
$^{21}${Minin University, Ulyanova st. 1, Nizhny Novgorod 603002, Russia}\\
$^{22}${Ka-Dar/Astroverty, Nizhny Arkhyz, Karachay-Cherkessia 369167, Russia}\\
$^{23}${Vedrus Observatory, Azovskaya, Krasnodar Krai 353245, Russia}\\
$^{24}${Crimean Astrophysical Observatory RAS, Nauchnyi 298409, Crimea}\\
$^{25}${Department of Astronomy, University of Illinois at Urbana-Champaign, 1002 W. Green Street, Urbana, IL 61801, USA}\\
$^{26}${Sternberg Astronomical Institute, Moscow State University, Universitetskii pr. 13, Moscow 119992, Russia}\\
$^{27}${Trevinca Skies, A Veiga，Ourense, 32360，Spain}\\
$^{28}${University of Arizona, Tucson, AZ 85721-0240, USA}\\
$^{\dagger}${These authors contributed equally to this work.}\\
$^{\ddagger}${Bengier-Winslow-Robertson Postdoctoral Fellow}\\
$*${Correspondence author: wang\_xf@mail.tsinghua.edu.cn}
}

\begin{document}
\maketitle

\noindent Shock breakout emission is light that arises when a shockwave, generated by core-collapse explosion of a massive star, passes through its outer envelope. Hitherto, the earliest detection of such a signal was at several hours after the explosion\cite{2018Natur.554..497B}, though a few others had been reported\cite{2008Natur.453..469S,2008Sci...321..223S, 2014Natur.509..471G, 2016ApJ...820...23G, 2017NatPh..13..510Y, 2022Natur.611..256C}. The temporal evolution of early light curves should reveal insights into the shock propagation, including explosion asymmetry and environment in the vicinity, but this has been hampered by the lack of multiwavelength observations. Here we report the instant multiband observations of a type II supernova (SN 2023ixf) in the galaxy M101 (at a distance of 6.85$\pm$0.15 Mpc\cite{2022ApJ...934L...7R}), beginning at $\sim$1.4 hours after the explosion. The exploding star was a red supergiant with a radius of about 440 solar radii. The light curves evolved rapidly, on timescales of 1$-$2 hours, and appeared unusually fainter and redder than predicted by models\cite{2010ApJ...725..904N, 2011ApJ...728...63R, 2023MNRAS.522.2764M} within the first few hours, which we attribute to an optically thick dust shell before it was disrupted by the shockwave. We infer 
that the breakout and perhaps the distribution of the surrounding dust were not spherically symmetric.

\medskip
\section{Main}~\label{sec:intro}
\begin{figure}
    \centering
    \includegraphics[trim={0.0cm 0.3cm 0.0cm 0.3cm},clip,width=1.0\textwidth]{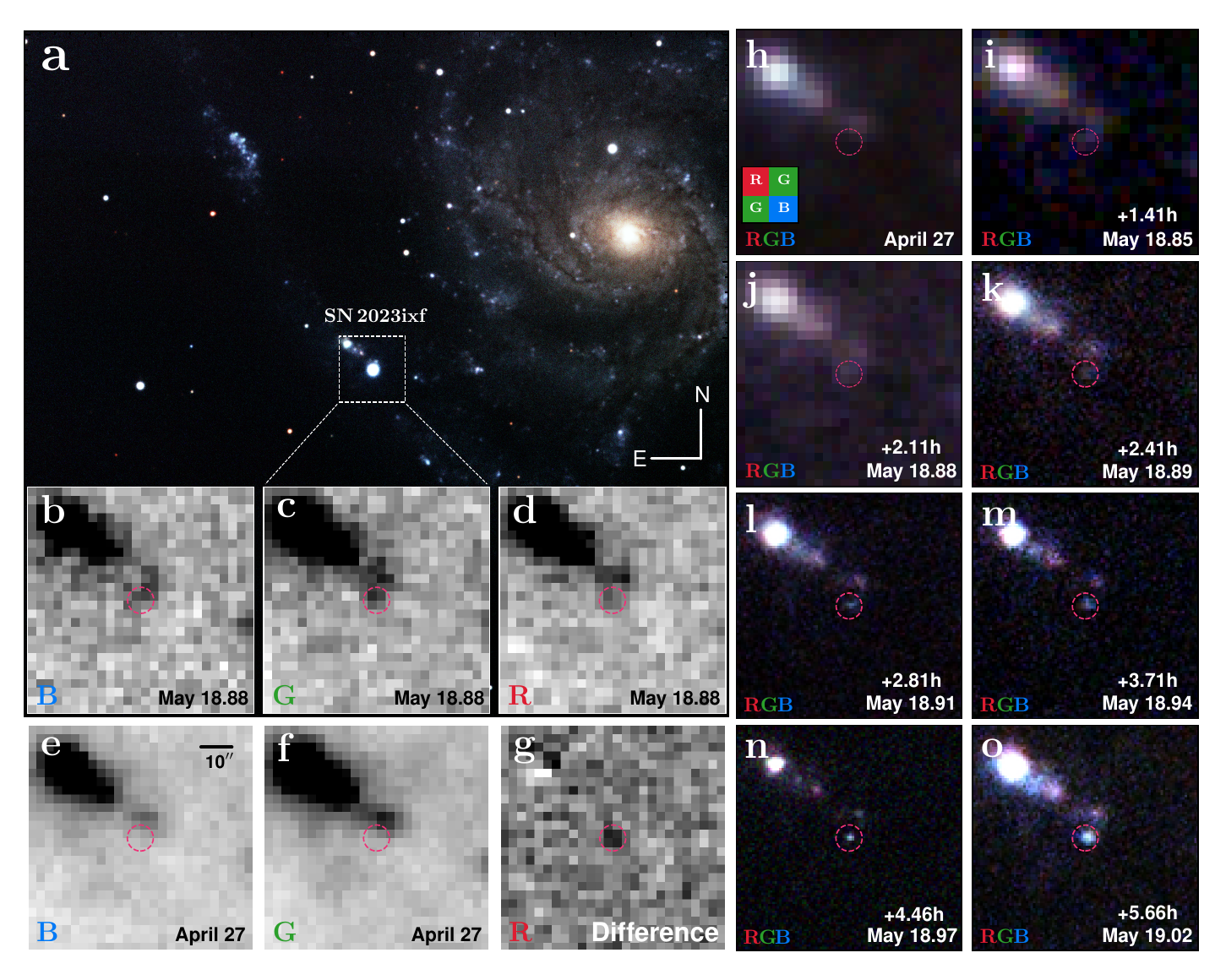}
    \vspace{-0.4 cm}
    \caption{The emergence of the first light of SN\,2023ixf.
    a: The upper-left panel shows a Johnson-Cousins $BVRI$-band composite image of SN\,2023ixf, obtained between UT 2023 May 20 and June 5 using YAHPT (see Method~\ref{sec:obs}). 
    b-d: A set of B (blue)-, G (green, middle)-, and R (red, right)-band images acquired by Xingming Observatory on 2023 May 18.88 to 18.89. The image set is generated by demosaicing the exposure through a Bayer filter (see Method~\ref{sec:amateur} and a schematic plot of the RGGB filter setup inlayed in panel h). e-g: The first two subpanels show the Bayer B- and G-band reference images obtained on 2023 April 27. The third subpanel presents an example of the Bayer R-band difference image. 
 h-o: A temporal sequence of RGB composite images centered at SN\,2023ixf, with the phase relative to the first light t$_{0}$ = 60,082.788 (May 18.788). 
 The red circle with a radius of 4$''$.2 marks the SN position.
   North is up and east is left.}
    \label{fig:discovery}
\end{figure}

The close distance makes SN\,2023ixf a nearby, bright stellar explosion that appears once in a decade, providing a rare apportunity to view the moment of shock breakout from the progenitor star. This SN was discovered on 2023 May 19.728 (UT dates are used throughout) by Koichi Itagaki, with an unfiltered CCD magnitude of 14.9 mag\cite{2023TNSTR1158....1I}. A spectrum taken at $\sim$5 hours later showed a series of prominent flash ionization lines of H, He, C, and N, suggestive of a type II supernova (SN II) explosion in its flash ionization phase\cite{2023TNSAN.119....1P}. Thanks to the efforts by amateur astronomers, the earliest detection of this SN can be traced back to 2023 May 18.829 \cite{2023TNSAN.130....1M}, about 0.9 days before the discovery. Photometric observations of SN 2023ixf from 10 different telescopes, including pre-discovery and instant followup observations, have been carefully compiled in Sloan $gr$ and Johnson $V$ bands (see Method~\ref{sec:amateur} and ~\ref{sec:obs}). In particular, the color images obtained through the Bayer RGB filters allow us to extract very early-time photometry in $B$, $G$ and $R$ bands, which are carefully calibrated to Sloan $g$, Johnson $V$, and Sloan $r$ bands, respectively (see Method~\ref{sec:amateur}). 
The early-color evolution unambiguously shows the emission evolved rapidly from red light at t$\sim$1-2 hours to blue light at t$\sim$3-5 hours after the explosion (see Fig.~\ref{fig:discovery}).
This color information is crucially important for the study of the properties of shock breakout emission and the opacity of the circumstellar dust. 

\begin{figure}
    \centering
    \includegraphics[width=0.98\textwidth]{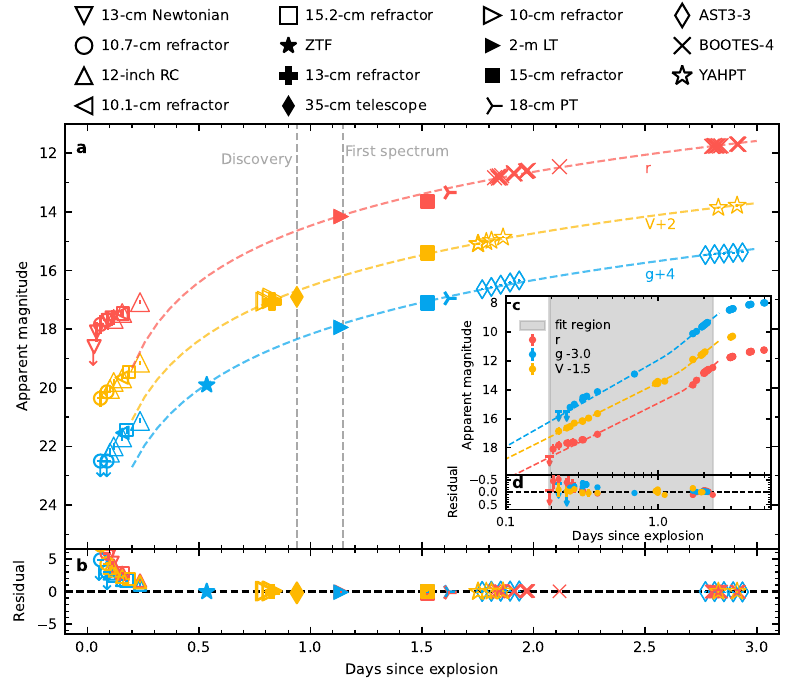}
    \caption{Optical light curves of SN\,2023ixf taken within the first 3 days relative to the estimated first light. Different symbols represent various facilities that are used to obtain the photometry (see top legend) in panel a. 
    Data points with downward arrows represent the non-detection limits, obtained within 16 minutes before the earliest detection in $r$ and $\sim$2.1 hours in $g$ (see Table ~\ref{tab:photometry}). The $g$- and $r$-band photometry on MJD 60,083.923 from the Liverpool Telescope (LT)\cite{2023TNSAN.119....1P} is also shown here but excluded in the fitting due to the lack of uncertainties. The epochs of the discovery and first spectrum are indicated with black dashed lines. 
    The colored dashed lines represent single power-law fits $f\propto (t-t_{0})^{n}$ $(n \approx 2.2)$ to the light curves taken before day +3. As illustrated by the large residuals displayed in panel b, the early photometric evolution of SN\,2023ixf exhibits significant deviations from an oversimplified expanding fire-ball model within $\sim$day 1. Panel c presents the best broken power-law fit to the early $g$- (blue dashed line), $V$- (yellow dashed line), and $r$-band (red dashed line) light curves. The vertical gray-shaded area indicates the time interval adopted for the light-curve fitting. The residuals are shown in panel d. The error bars shown represent 1-$\sigma$ uncertainties of magnitudes and phases.
    }
    \label{fig:fireball}
\end{figure}

Fig.~\ref{fig:fireball} shows the multiband light curves collected within 3.0 days after the first detection (see Table ~\ref{tab:photometry} for detailed photometry). In the first few hours of the initial detection, the emission of SN\,2023ixf appears stronger than that predicted from the fire-ball model by $\sim$100 times. This large discrepancy indicates that additional significant energy contributes to the early-time radiation in optical bands, which can be only caused by emission following the shock breakout at such an early phase. Owing to a complex luminosity evolution in the early phase, we applied a broken-power law fit to the first $\sim$2.5-day $gVr$-band photometry to estimate the first-light time (see Method~~\ref{sec:first_light}). Despite relatively large photometric errors, the fitting to the first few-hour data appears not satisfactory, especially in the well-observed $r$ band. A hybrid model with considerations of multiple processes provides a plausible fit to the pre-peak evolution following shock breakout emission
(see Method~\ref{sec:hybrid}). The inferred explosion time t$_{0}$ (MJD 60082.788$^{+0.02}_{-0.05}$) is adopted as the reference time throughout this paper. With this t$_{0}$, the earliest flash recorded for SN 2023ixf occurred at t$\sim$0.98$^{+1.2}_{-0.48}$ hour after explosion, while the color information is available at about 0.4 hour later.

\begin{figure}
     \centering
    \includegraphics[width=0.98\textwidth]{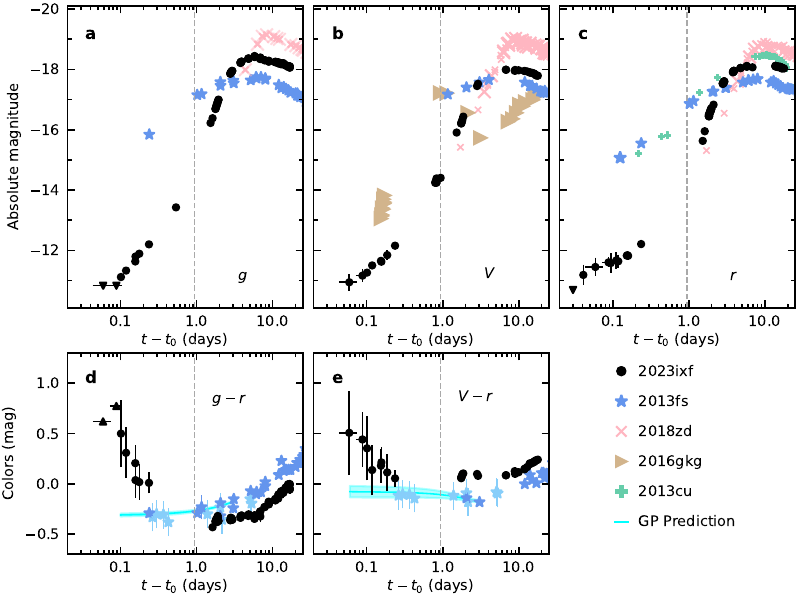}
    \caption{Temporal evolution of the early brightness and colors of SN\,2023ixf.
    a-c: Multi-band photometric evolution of SN\,2023ixf compared with that of a selected sample of SNe II that exhibit prominent signatures of shock-cooling tail (SN\,2016gkg, \cite{2017ApJ...837L...2A}) or short-lived `flash' features in the early phase of explosion, namely SNe\,2013cu\cite{2014Natur.509..471G}, 2013fs\cite{2017NatPh..13..510Y}, and 2018zd\cite{2020MNRAS.498...84Z}. The above comparison SNe have a rather complete coverage of the rising phase. All phases are relative to the estimated time of the first light of each SN. d and e: Comparison of the early-time $g-r$ and $V-r$ color evolution between SN\,2023ixf and SN 2013fs\cite{2017NatPh..13..510Y}. The light blue stars represent the synthetic colors calculated from the early spectra of SN\,2013fs. 
    The cyan segments present the color trends extrapolated from SN\,2013fs using the Gaussian Process (GP) regression to  $\sim$1.4 h after explosion, with the shaded areas indicating 1-$\sigma$ uncertainties. The error bars shown represent 1-$\sigma$ uncertainties of magnitudes, colors, and phases.
        All magnitudes and colors displayed have been corrected for the Galactic and the host-galaxy extinctions (see Method \ref{sec:comparison}).  The discovery time is marked with a vertical grey dashed line.
    }
    \label{fig:r_comparison}
\end{figure}

In Fig.~\ref{fig:r_comparison}, the early photometric evolution of SN\,2023ixf is compared with that of a few well-observed SNe II at early phases (see Method~\ref{sec:comparison}). Though having similar near-peak photometric evolution, SN\,2023ixf stands out relative to the comparison ones for its rapid rise and remarkable flux deficit within the first 2 days. At t$\sim$5 hours, SN\,2023ixf appears fainter by $\sim$4.0 mag in $g$ and $\sim$3.0 mag in $r$, respectively. Additionally, a red excess is clearly seen in the color curves within the first $\sim$5 hours, with differences of $\sim$3–4$\sigma$ in $g-r$ and $\sim$2$\sigma$ in $V-r$ relative to SN 2013fs
\cite{2017NatPh..13..510Y}
(see Fig.~\ref{fig:r_comparison} d and e). Its origin will be further explored below through semi-analytic modeling of shock breakout emission. From t$\sim$ 5 hours post-explosion, the color evolution of SN\,2023ixf becomes overall consistent with that of the comparison SNe II\cite{2014Natur.509..471G,2017NatPh..13..510Y}, suggesting a similar shock-cooling temperature afterwards.  

For core-collapse supernovae in the early phase, their emission from the rapidly expanding and adiabatically cooling ejecta dominates over the radioactive heating from the innermost regions of the explosion \cite{2010ApJ...725..904N, 2011ApJ...728...63R, 2014ApJ...788..193N, 2017ApJ...838..130S}. In comparison with other SNe II, 
the emission of SN 2023ixf increased sharply over $t\sim+1$ to $+$5 days, suggesting a significant contribution by the interaction of SN ejecta with circumstellar material (CSM). The narrow and short-lived highly-ionized line species that persist over similar phases \cite{2023TNSAN.119....1P, 2023ApJ...954L..42J,zhang2023circumstellar} also imply an interaction process within any surrounding dense CSM (with a typical number density $n \sim 10^{-12}$ cm$^{-3}$ and a confined radial distribution within $\sim10^{13}-10^{15}$ cm, \cite{2017NatPh..13..510Y}). Therefore, we propose a hybrid analytic model (see Method~\ref{sec:hybrid}) to explain the photometric evolution of SN\,2023ixf by considering the emission due to shock breakout and that from the CSM interaction. For the shock breakout component, we adopt a semi-analytic solution \cite{2010ApJ...725..904N} to fit the $g$-, $r$-, and $V$-band light curves of SN\,2023ixf at early times. Owing to an enhanced (and potentially eruptive) mass loss when approaching the core collapse, we employ a broken power law radial CSM density profile (see Method~\ref{sec:interaction}), i.e., a dense shell confined within a distance of $10^{14}-10^{15}$\,cm from the progenitor \cite{2023ApJ...956...46S}. This dust shell could be formed through a huge rate of pre-explosion mass loss, i.e., $\sim10^{-3}-10^{-2}$\,M$_{\odot}$\,yr$^{-1}$ within years before the final explosion \cite{2023ApJ...954L..42J, 2023ApJ...956...46S, 2023ApJ...956L...5B, 2023ApJ...954L..12T,zhang2023circumstellar}. The best fit to the multi-band fast-evolving optical light curves of SN\,2023ixf suggests that the CSM shell spans a range of $4.0$-$6.0\times10^{14}$ cm, corresponding to a mass-loss rate of $5.0\times10^{-2}$ M$_{\odot}$ yr$^{-1}$ (see Table ~\ref{tab:interaction}), consistent with the configuration obtained by other group\cite{2023ApJ...955L...8H}. 

\begin{figure}
    \centering
    \includegraphics[width=0.98\textwidth]{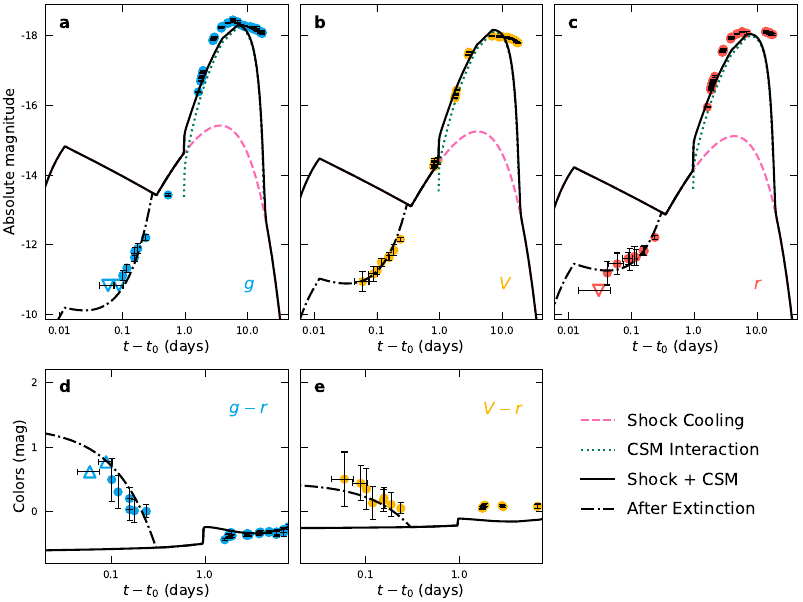}
    \caption{Fitting the shock breakout emission of SN 2023ixf with a multiple-process hybrid model.
    a-c: Best-fit models to describe the early $g$-, $V$-, and $r$-band flux evolution, respectively. For each band, dashed pink, dotted green, and solid black lines represent the fits to the shock-cooling, the ejecta$-$CSM interaction, and their combined flux, respectively. 
    Open triangles indicate non-detection limits. The effect of the time-variant extinction term, which is essential to account for the significant flux depression within the first $\sim$5 hours after the estimated time of the explosion, is shown by the dotted-dashed black line. 
    d and e: Similar to the upper panels but for the $g-r$ and $V-r$ color evolution, respectively. Only the fit to the combined emission from both the shock cooling and the ejecta$-$CSM interaction is presented, which is shown by the black solid line in each panel.  The error bars indicate 1-$\sigma$ uncertainties of magnitudes, colors, and phases.
    }
    \label{fig:modelfit}
\end{figure}

Apart from the first few hours after the explosion, our shock cooling + ejecta$-$CSM interaction model produces a reasonable fit to the pre-peak multi-band light curves of SN\,2023ixf, as shown in Fig.~\ref{fig:modelfit}. Our estimated mass-loss rate and the location of the dense CSM shell are in very good agreement with those derived from the hydrodynamic simulations with time-series spectroscopy of SN\,2023ixf \cite{2023ApJ...954L..42J,2023ApJ...956L...5B,2023ApJ...954L..12T,2023ApJ...956...46S}. The fitting is also consistent with the limits placed by the millimeter observations\cite{2023ApJ...951L..31B}.

To account for the significant flux deficit in blue band and unusual color evolution within the first few hours after the explosion, we consider a time-variant extinction term for SN\,2023ixf. Such an extinction variation can be caused by the sublimation of line-of-sight (LOS) dust grains in the CSM. In our modeling, we assume an initial optical depth that decreases monotonically and only persists within the first $\sim$0.3 days. 
The introduction of a linearly-decreased optical depth can significantly improve the fit to the multi-band luminosity and color evolution observed during the very early phase (see Fig.~\ref{fig:modelfit}). One natural interpretation gives the photo-destruction of primordial dust grains surrounding the exploding star by the initial UV-Optical flash, which has been discussed and observed in a few Gamma-Ray Bursts\cite{2000ApJ...537..796W, 2014MNRAS.440.1810M}. For dust grains with a radius size of $a$, the survival time at temperature $T$ can be estimated as 
$t_{\rm surv}$ = 7.7$\frac{a}{0.1\,{\rm \mu m}}$ exp$[7 \times 10^{4}\,{\rm K}(\frac{1}{T} - \frac{1}{\rm 2300\,K})]$\,s 
\cite{2000ApJ...537..796W}. 
Within the first few hours after $t_{0}$, the radiation $T$ would be at least a few tens of thousands K as indicated by the presence of highly-ionized flash features. This is much higher than the typical sublimation temperature of $\sim$2000\,K for carbonaceous, graphite, and silicate grains\cite{2001ApJS..134..263W}, thus leading to an immediate thermal sublimation of the CSM dust by the high energy shock photons.
 
For the CSM confined within $\sim5\times10^{14}$\,cm from SN\,2023ixf \cite{2023ApJ...956...46S}, if the breakout emerges isotropically, all dust grains will be immediately sublimated when the first light is received by the observer. Thus, an aspherical shock breakout is needed to prolong the duration of the shock front and make the dust grains survive longer along the directions with larger optical depth. 
The initial breakout front may wrap around the stellar surface starting from where the breakout acceleration initially takes place \cite{2021MNRAS.508.5766I}. 
Similar effects from such a prolonged breakout are also suggested by recent 3D hydrodynamic calculations\cite{2022ApJ...933..164G}. 
To the first order, such an aspherical breakout would result in a complex interplay between multiple time-varying emission sources and sublimation of dust grains within $\sim$hour timescales (see Method~\ref{sec:aspherical} for detailed discussions). Reconciling this interplay requires more detailed modeling in the future.

\begin{figure}
    \centering
    \includegraphics[width=0.8\textwidth]{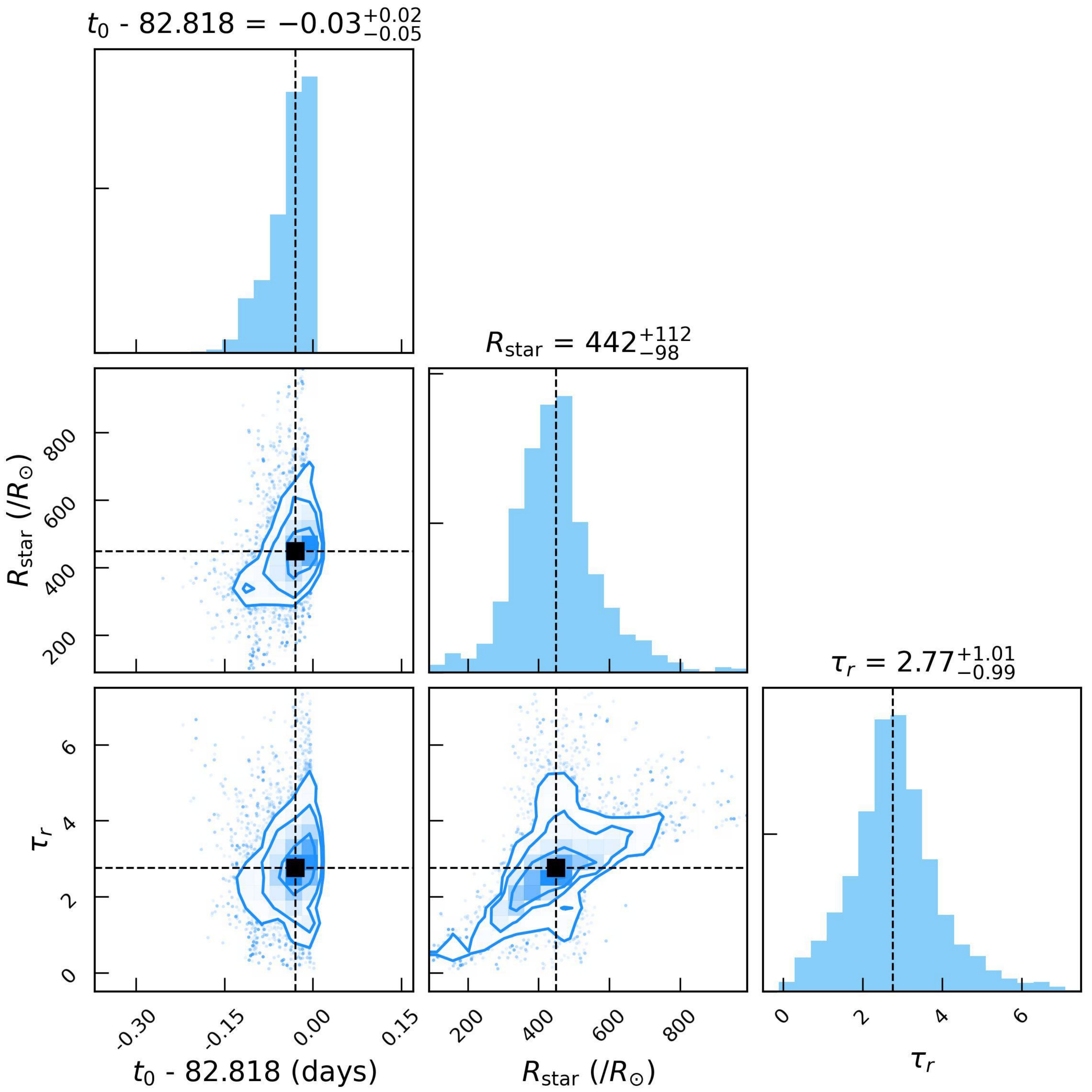}
    \caption{Joint confidence level contours of the parameters inferred from the MCMC-based fitting. The demonstrated parameters are the time of the first light $t_{0}$ (in a unit of days after MJD 60,000.0), the radius of the progenitor star $R_{\rm star}$ (in a unit of solar radius), and the $r$-band optical depth of the circumstellar dust measured at $t_{0}$. Best-fit parameters are marked by horizontal and vertical lines and labeled with their 1$\sigma$ confidence levels. The inner, middle, and outer contours centered at any intersection of parameter pairs show the 68\%, 95\%, and 99.7\% confidence levels, respectively. The likelihood histograms are scaled to 1 for all the three parameters.
}
    \label{fig:modelfit_contour}
\end{figure}

The best-fit parameters of our hybrid model (see  Fig.~\ref{fig:modelfit_contour}) indicate that we captured the first-hour shock flash of SN 2023ixf after the explosion, representing the earliest electromagnetic signal ever observed for a core-collapse SN. Its progenitor is estimated to have a radius $R_{\rm star}=$442$^{+112}_{-98}$ R$_{\odot}$, favoring a red supergiant origin, which is consistent with recent analysis\cite{2023ApJ...957...64S,2023ApJ...953L..16H,2023ApJ...952L..23K,2023arXiv230901389X}. The inferred significant initial LOS extinction due to the circumstellar dust, i.e., $\tau_{r}$= 2.77$^{+1.01}_{-0.99}$, which corresponds to a $V$-band extinction A$_{V}\sim$3.5 mag, indicates the presence of an optically-thick CSM shell that is likely to have caused an aspherical shock breakout. The very early chromatic movie recorded for SN\,2023ixf also hints that within a few years before death its progenitor may have experienced eruptive or pulsational mass loss, which was likely intense, dusty and inhomogeneous.

\clearpage

\clearpage

\clearpage
{\hspace{-0.3em}}
\section{Method}

\subsection{Photometry of pre-discovery observations}~\label{sec:amateur}
Pre-discovery observations were obtained from multiple amateur astronomers using various facilities. These include: a 10.7-cm refractor equipped with ZWO ASI071MC-Cool camera with a Bayer RGGB filter, observed at Dabancheng, Xinjiang, China \cite{2023TNSAN.130....1M}; a 13-cm Newtonian telescope equipped with unfiltered ZWO ASI1600MM Pro monoch-rome CMOS camera 
\cite{2023TNSAN.150....1C}; a 10.1-cm refractor equipped with QHY600m CMOS with RGB filter wheel observed in Russia\cite{2023TNSAN.150....1C}; a 15.2-cm refractor equipped with ZWO ASI6200MM Pro camera with RGB filters observed at Sola, Norway; a 12-inch Ritchey–Chr\'etien (RC) telescope equipped with ATR3CMOS 26000KPA with a Bayer RGGB filter, observed at Trevinca, Spain\cite{2023TNSAN.127....1H}; a 0.6-m reflector equipped with FLI 230-42 (NEXT) with Sloan filters observed at Xingming Observatory in Xinjiang, China; a 13-cm refractor equipped with ZWO ASI2600MM Pro with a Luminance(L)filter, observed in Yunnan, China \cite{2023TNSAN.130....1M}; a 10-cm refractor equipped with ZWO ASI2600MM with a L-band filter, observed in Yunnan, China\cite{2023TNSAN.130....1M}; and a 15-cm refractor equipped with FLI Proline 10002M with RGB filter observed in Utah, the USA\footnote{Observed after discovery but has been reduced using the same method as the other images in this section}. Bias subtraction, dark current correction, and flat-field correction of those images were performed with \textsc{Astropy} package \cite{2013A&A...558A..33A,2018AJ....156..123A}.

The images captured with Bayer filters were decomposed into Red (R), Green (G), and Blue (B) channels using the \textsc{Astropy} package. The RGGB Bayer pattern is commonly used in digital image sensors and is characterized by a recurrent 2 $\times$ 2 array arrangement. Within each of these arrays, there are two green pixels, one red pixel, and another blue pixel. A visual representation of this pattern is displayed in Fig.~\ref{fig:discovery}h. The data of each color are restructured with a half dimension of original images (given the 2 $\times$ 2 Bayer pattern). This results in four distinct images representing the R, G1, G2, and B channels, respectively. The header information of the original FITS files was adapted for each of the extracted channel images. Specifically, the dimensions were halved, and the filtered metadata was updated to reflect the respective channel. Such an extraction process enables a breakdown of color information. 

Stacking of images was performed to improve the signal-to-noise ratio (SNR) for the 10.7-cm refractor data, using the \textsc{reproject} package\footnote{https://reproject.readthedocs.io/en/stable/index.html}. Each stacked image has a total exposure time of 45 min. Similarly, the images from the 13-cm Newtonian telescope were stacked with exposure times of 13.3 min, 26.7 min, and 20 min, separately, resulting in photometry on May 18.83, 18.88, and 18.90, respectively.

For the RGB images obtained by the 10.7-cm refractor and the 15.2-cm refractor, template subtraction was performed using reference images acquired prior to SN explosion with the same observation setup. This subtraction was done using the \textsc{PyZOGY} package\footnote{https://github.com/dguevel/PyZOGY}\cite{2016ApJ...830...27Z}.
Aperture photometry was carried out on the subtracted images of the 10.7-cm refractor using \textsc{AUTOPHOT} \cite{AUTOPHOT}, considering its large pixel scale (2$''$.8 pixel$^{-1}$) and relatively low signal-to-noise ratio (SNR) of SN\,2023ixf. Aperture photometry was also performed on the images of the 15.2-cm refractor due to the faintness of the SN. For the images obtained by the other telescopes, point spread function (PSF) photometry was conducted using \textsc{AUTOPHOT}.

\begin{figure}
    \centering
    \includegraphics[width=0.7\textwidth]{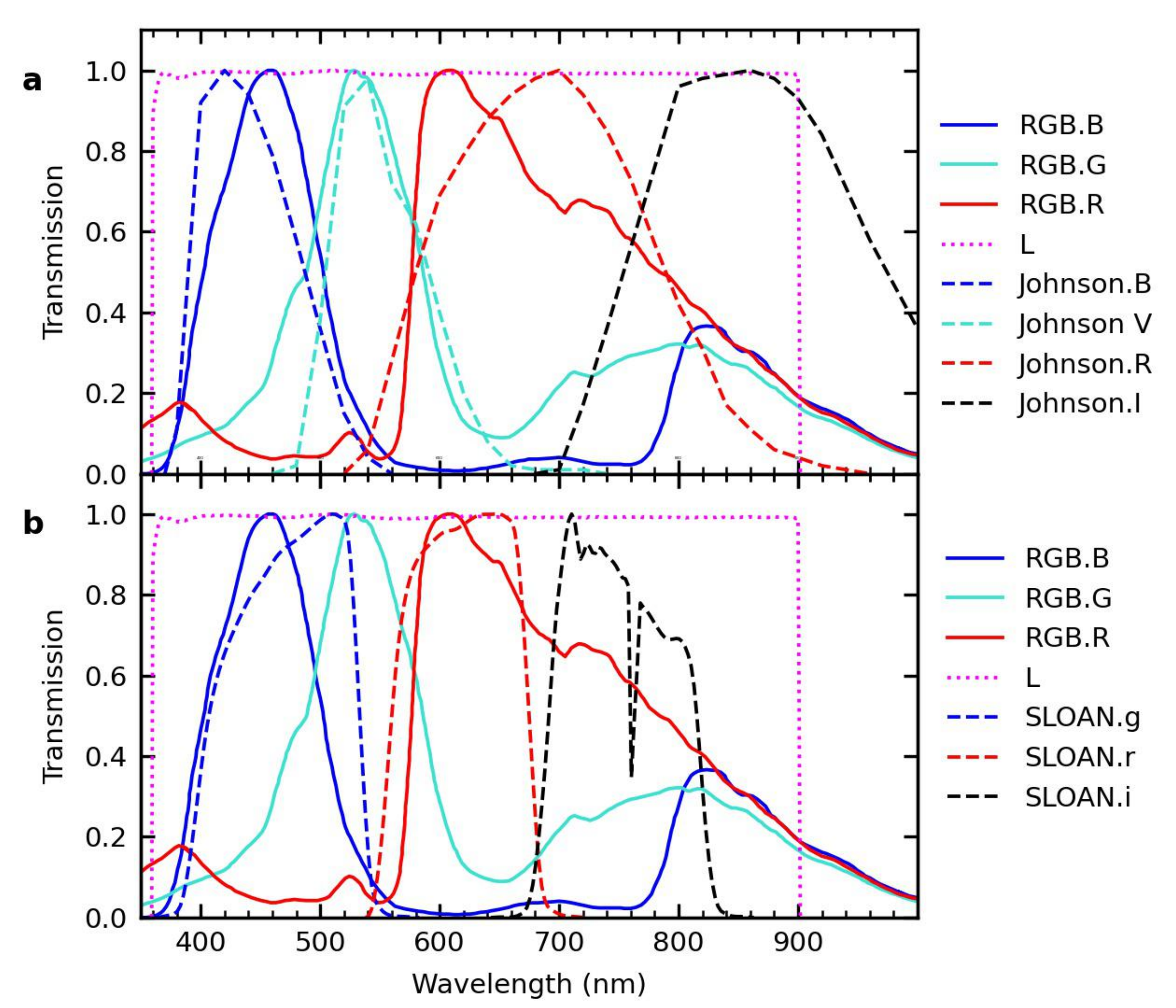}
    \caption{Transmission curves of different photometric systems. Comparison of the transmission curves of the RGB filters with the standard Johnson/Bessell $BVRI$ (a) and Sloan $gri$ filters (b) used in the observations of SN\,2023ixf presented in this paper. The transmission curves are normalized to the peak transmission rate for each filter.
}
    \label{fig:filter_curve}
\end{figure}

\begin{figure}
    \centering
    \includegraphics[width=0.7\textwidth]{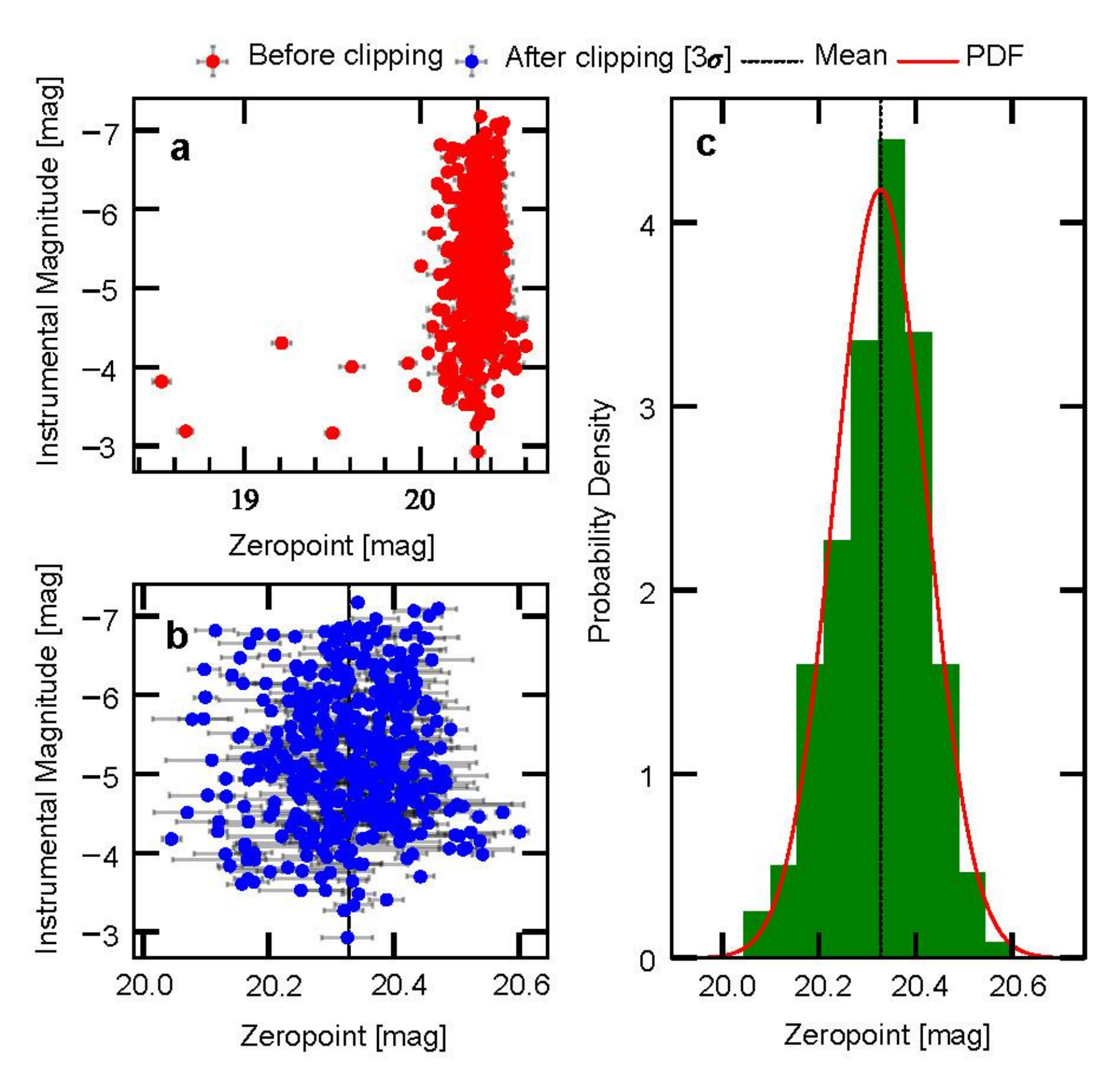}
    \caption{Zero-point diagnostic plot generated by \textsc{AUTOPHOT}. Zero-point measurements before (a) and after (b) a 3$\sigma$ clipping process. The zero-point value is indicated by a vertical solid line. The error bars indicate 1-$\sigma$ uncertainties of zeropoints. Panel (c)  displays the probability density function of the zero-point distribution with a well-defined peak.}
    \label{fig:zp}
\end{figure}

To reliably calibrate the zeropoints of the R-, G-, B-, L-band, as well as unfiltered images, we compared the transmission curves of the filters with those of the Sloan Digital Sky Survey (SDSS) and Johnson-Cousins broadband photometric systems\footnote{The RGB transmission curves are retrieved from https://zwoasi.com/manuals/ASI071MC\_Pro\_Manual\_CN\_V1.5.pdf} (see Fig.~\ref{fig:filter_curve}). We chose to calibrate the zero-points of RGB R to SLOAN r band, RGB G to Johnson V band, RGB B to SLOAN g band, L to Johnson V band, and the unfiltered images to SLOAN r band, respectively. For the convenience of comparison, the zeropoint of RGB B-band are also calibrated to that of the Johnson $B$-band, as listed in Table ~\ref{tab:photometry}. All images were flux calibrated against the APASS catalog \cite{2015AAS...22533616H}. To derive the zero points, we measured the instrumental magnitude ($m_{inst}$) of point sources in each image. Sources with an SNR $>$ 50 were selected, which were subsequently compared with their catalog magnitudes in the corresponding band ($m_{cat}$). The zero point (ZP) for each source was computed using the relation $m_{ZP}$ = $m_{cat} - m_{inst}$. To ensure the robustness of our measurements, an iterative 3-$\sigma$ clipping method was employed to omit the outliers of the ZP. We then engaged \textsc{LMFIT} library \footnote{https://lmfit.github.io/lmfit-py/} \cite{2014zndo.....11813N} to execute a vertical line fit between $m_{cat}$ and $m_{ZP}$ (see Fig.~\ref{fig:zp}a,b), furnishing us with the image ZP and the associated standard error. The typical uncertainty of ZP calibration is less than 0.1 mag (see Fig.~\ref{fig:zp}c). The photometry results are listed in Table ~\ref{tab:photometry}.

\begin{figure}
    \centering
    \includegraphics[width=0.9\textwidth]{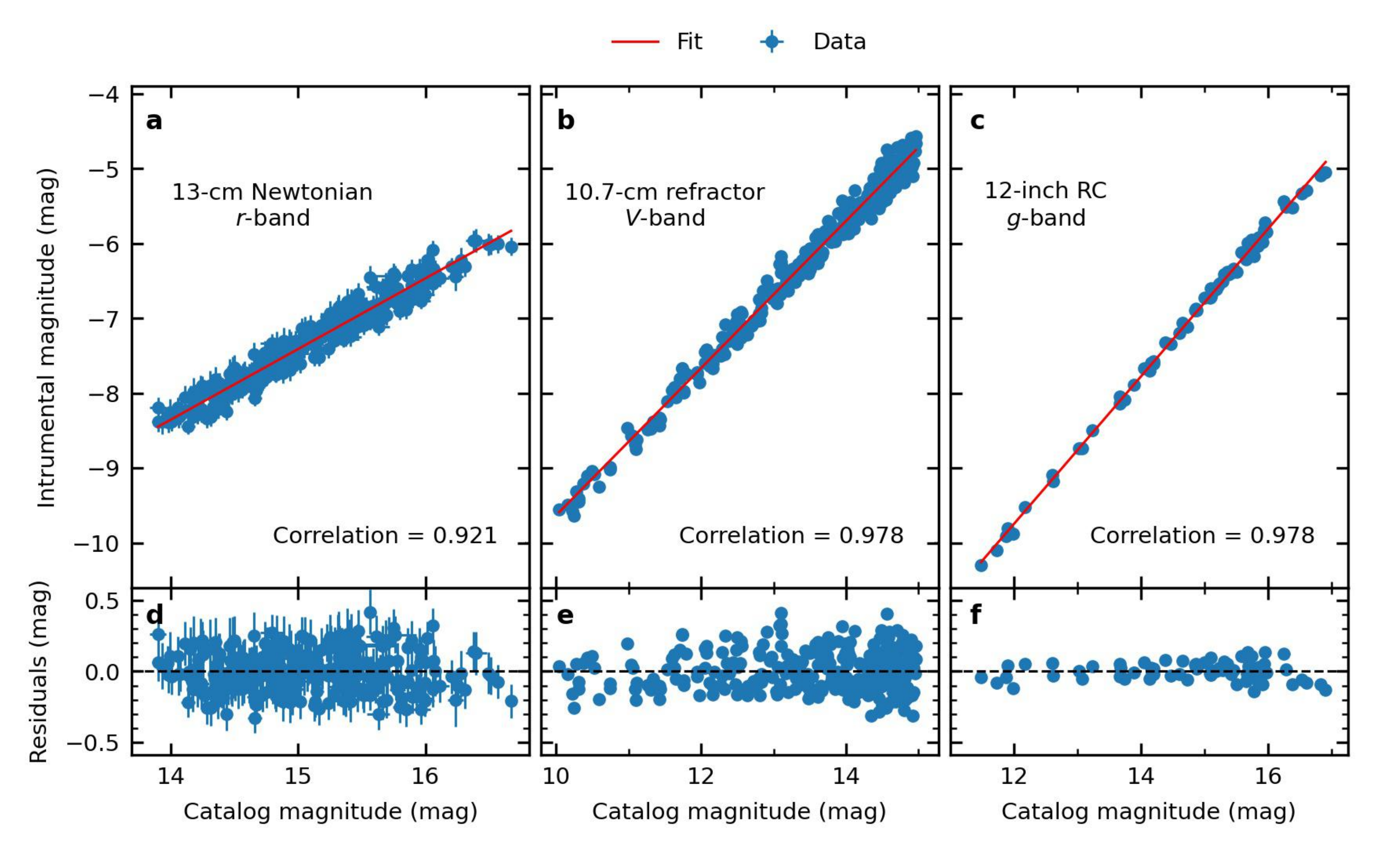}
    \caption{Assessment of the instrument linearity. Panels (a-c) display a scatter plot that compares the instrumental magnitudes against the catalog magnitudes for the 13-cm Newtonian, 10.7-cm refractor, and 12-inch RC telescopes, respectively. The red lines denote the best-fit linear regressions. Panels (d-f) display the residuals from the regressions. A horizontal red dashed line at zero serves as a reference. The error bars indicate 1-$\sigma$ uncertainties of instrumental and catalog magnitudes.}
\label{fig:linear}
\end{figure}

To assess the linearity of those instruments, we conducted a linear regression analysis for $m_{cat}$ and $m_{inst}$ using the \textsc{linregress} function from the \textsc{SciPy} package\cite{2020NatMe..17..261V}.
For each instrument, we computed the residual between $m_{inst}$ and its corresponding prediction from the regression model.
These residuals aided in the identification of outliers, deploying the interquartile range (IQR) as the criterion. Data points with residuals deviating the bounds set by \( Q1 - 1.5 \times \text{IQR} \) and \( Q3 + 1.5 \times \text{IQR} \) were marked as outliers and excluded from the dataset. The \( Q1 \) and \( Q3 \) denote the first and third quartiles of the residuals, respectively, and the IQR is defined as \( Q3 - Q1\).  Such a procedure was iteratively executed to remove outliers from the dataset, and this approach was applied to all of our data. The assessment on the images of three main instruments is illustrated in Fig.~\ref{fig:linear}, with the Pearson coefficients ($corr$) of 0.90-0.98 for $m_{cat}$ and $m_{inst}$, suggesting a satisfactory linearity for our photometry.

\subsection{Photometry of followup observations}~\label{sec:obs}
 
 The photometric follow-up campaign on SN\,2023ixf has been carried out promptly by the 0.8~m YaoAn High Precision Telescope (YAHPT), Antarctic Survey Telescope (AST3)-3 \cite{AST3_22Univ} at the Yaoan station of Purple Mountain Observatory, China, and the 0.6~m BOOTES-4 telescope\footnote{4th site of Burst Optical Observer and Transient Exploring System} at Lijiang station of Yunnan Observatories, China, the 0.8\,m Tsinghua University-NAOC\footnote{National Astronomical Observatories of China (NAOC), China.}  telescope (TNT)\cite{2008ApJ...675..626W}) at Xinglong Station of NAOC, China,
the 0.36\,m reflector of Tsinghua University (SNOVA) at Nanshan Station of Xinjiang Observatory, China.

All images were pre-processed following standard routines, including bias subtraction, flat-field correction, dark current correction, and cosmic ray removal. 
The $g$-band images of SN\,2023ixf taken by AST3-3 were processed using the pipeline described in \cite{AST3_22FrASS}. 
Archival $g$-band AST3-3 images obtained on June 2021 were adopted for background subtraction using the SFFT\footnote{https://github.com/thomasvrussell/sfft} algorithm \cite{SFFT_Hu2022}. Aperture photometry has been performed on differential images using SourceExtractor \cite{SExtractor} and calibrated to the APASS DR9 catalog\cite{2016yCat.2336....0H}. 
YAHPT images were processed with the CCDPROC \cite{matt_craig_2017_1069648_ccdproc} package and solved for astrometry using astrometry.net \cite{2010AJ....139.1782L_astrometry.net}, with respect to the Gaia DR2 catalogue \cite{2016A&A...595A...1G, 2018A&A...616A...1G}. Aperture photometry has also been performed on YAHPT images with Source Extractor. 
Aperture/PSF photometry has been also performed for the BOOTES-4, SNOVA and TNT images, with the instrumental magnitudes calibrated using the APASS catalog\cite{2015AAS...22533616H}.
All of the early photometric data used for model fitting are displayed in Table ~\ref{tab:photometry} and Fig.~\ref{fig:fireball}. 
In addition, the $g$- and $r$-band photometry of MJD 60,084.407 from the Plaskett Telescope (PT)\cite{2023TNSAN.129....1K} and the $g$-band photometry of MJD 60,083.323 from the Zwicky Transient Facility (ZTF)\cite{2023TNSAN.120....1P} are included in our model fitting.

\subsection{Early-time luminosity evolution and first-light time}~\label{sec:first_light}
\begin{figure}
    \centering
     \includegraphics[width=1.0\textwidth]{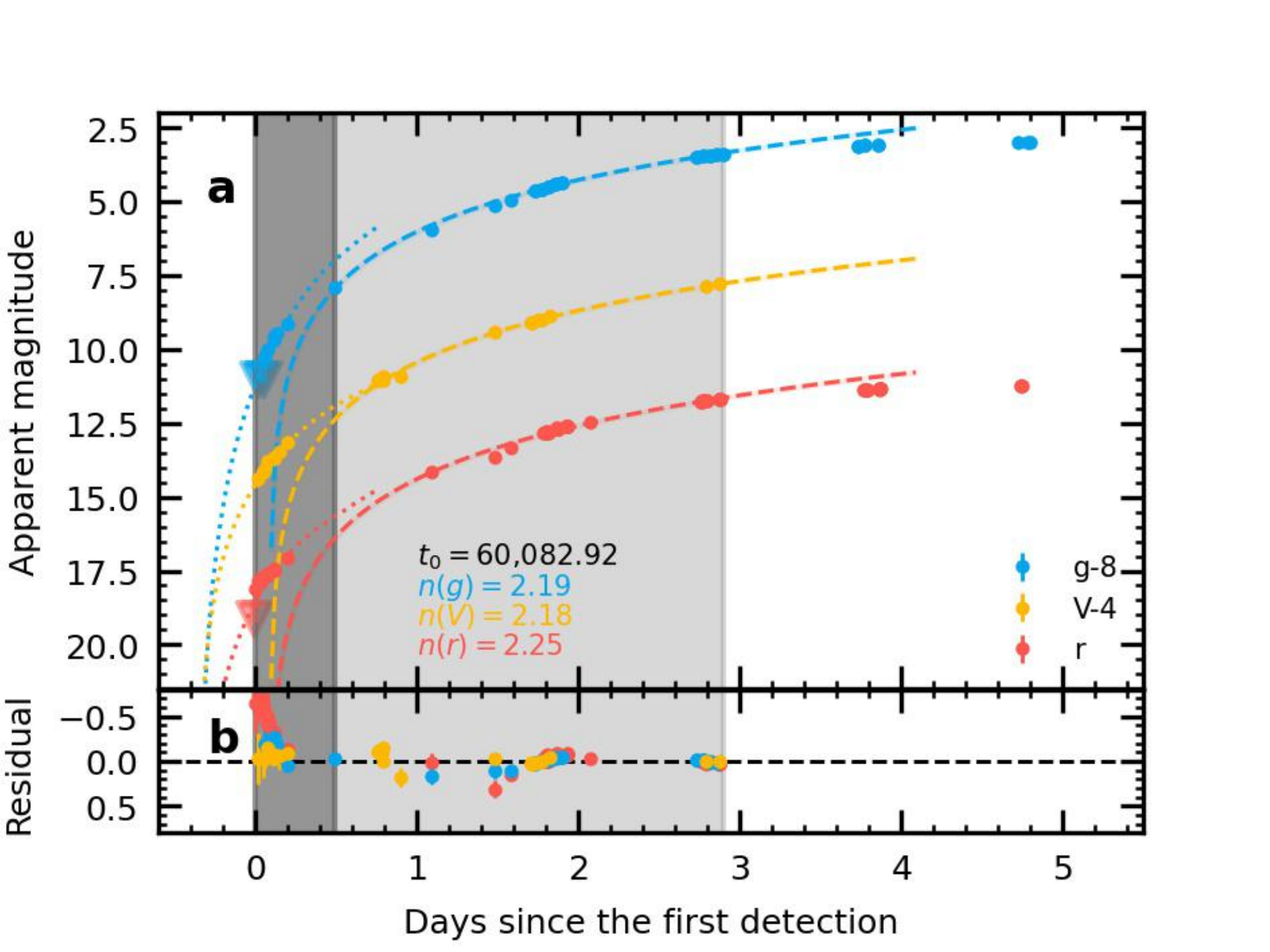}
     \caption{Broken power-law fit to the early-time photometry of SN\,2023ixf. (a): Separate fitting to the t $<$ 0.4 day (dark gray region) and 0.4 d $<$ t $<$ 3.0 day (gray region) phases of the early-time light curves of SN 2023ixf with $f\propto(t-t_{0})^n$ model, with the best fitting curves represented as dotted and dashed lines, respectively. The $g$- and $V$-band light curves are shifted vertically for better display. The shifted values are shown in the legends. The error bars indicate 1-$\sigma$ uncertainties of magnitudes. (b): The residuals relative to the best fittings.
}
     \label{fig:ftn_rgV0.5-3d_mcmc}
\end{figure}

Different models have been proposed to fit the early-time luminosity evolution of SN, i.e., fireball model expressed as $f\propto(t-t_{0})^2$ \cite{riess1999there}. For SN\,2023ixf, the flux at early times rises much faster than that predicted by the fireball model, we thus adopt a broken-power law model\cite{2013ApJ...778L..15Z} to fit the early data within $\sim$2.0 days after the first detection. The fitting formula is expressed as below 
\begin{align}
f(t) = A\left(\frac{t-t_0}{t_b}\right)^{\alpha_1}{\left[1+{(\frac{t-t_0}{t_b})}^{s\left(\alpha_1-\alpha_2\right)}\right]^{-1/s}}    
\end{align}
where $f(t)$ is the flux at time $t$, $A$ gives the scaling factor, $t_{0}$ and $t_{b}$ represent the first-light time and the time of the break, respectively. $\alpha_1$ and $\alpha_2$ denote the power-law indexes as fitted before and after $t_{b}$, respectively, while $s$ is a parameter introduced to describe the “smoothness” of the slope change. 
We applied a joint broken-power law fitting to the $g$-, $V$- and $r$-band light curves of SN 2023ixf. With a Markov chain Monte Carlo (MCMC) sampling approach, we find that $\alpha_{1}$ is systematically smaller than $\alpha_{2}$, and the fitting to the $r$-band data is unsatisfactory with a reduced $\chi^{2}$=4.6 (see Table ~\ref{tab:early_lc}). These results are consistent with that the SN experienced a complicated luminosity evolution at early time, especially at the first few hours after explosion. The joint fitting yields the first-light time as MJD  $60082.63^{+0.07}_{-0.05}$, however, this determination may be significantly biased as the earliest detections would be significantly affected by shock breakout emission and do not follow a simple power-law evolution\cite{2010ApJ...725..904N}. 
After excluding the earliest few data points (i.e., t $<$ 0.4 day), a single power-law joint fit to the rest data (i.e., 0.4 $<$t$<$ 3.0 day) in $grV$ bands yield t$_{0}$ as MJD 60082.92$^{+0.01}_{-0.01}$, while this time is later than the earliest detections and thus considered unreasonable (see Fig.~\ref{fig:ftn_rgV0.5-3d_mcmc}). Thus a hybrid shock cooling and interaction model is proposed to fit the early-time luminosity in Method~\ref{sec:hybrid}

\subsection{Parameters of the comparison sample}\label{sec:comparison}
The comparison sample shown in Fig.~\ref{fig:r_comparison} includes SNe\,2013fs (a normal SN IIP, \cite{2017NatPh..13..510Y}), 2018zd (a luminous and slowly-expanding SN IIP, \cite{2020MNRAS.498...84Z}), 2016gkg (an SN IIb that displayed a shock-cooling emission, \cite{2017ApJ...837L...2A, 2018Natur.554..497B}) and 2013cu (an SN II showing prominent flash spectral features, \cite{2014Natur.509..471G}). Among this sample, SN\,2016gkg is reported to have the earliest detection at t$\sim$4.3 hours after the explosion, while the earliest detections proposed for SNe\,2013cu and 2013fs are at t$\sim$5.7 hours and t$\sim$3.0 hours, respectively. Nevertheless, the explosion time of SN\,2013fs is inferred with a polynomial fit to the limited early-time data and may suffer large uncertainties. To compare with the early colors of SN\,2023ixf, we utilized Gaussian Process (GP) regression via the \textsc{george} library\footnote{https://github.com/dfm/george} to extrapolate the colors of SN\,2013fs to earlier phase.

For all SNe presented, reddening correction has been applied to all the light and color curves. For SN 2023ixf, a total reddening of $E(B-V)=0.04$ mag is adopted, including the galactic reddening of $E(B-V)=0.01$~mag\cite{2011ApJ...737..103S} and the host component of $E(B-V)=0.03$ mag inferred from the high resolution spectra of SN~2023ixf \cite{2023ApJ...956...46S,2023ApJ...954L..12T}. 
A total reddening of $E(B-V)=0.05$ mag and a distance of $D=50.95$ Mpc are adopted for SN\,2013fs\cite{2017NatPh..13..510Y}. Absolute magnitudes of SN\,2013cu are taken from reference\cite{2014Natur.509..471G}. For SN\,2018zd, we adopt a reddening of $E(B-V)=0.40$ mag and distance of $D=18.4\pm4.5$ Mpc\cite{2020MNRAS.498...84Z}. For SN\,2016gkg, the reddening is assumed as $E(B-V)=0.017$ mag and the distance is adopted as $D=26.4$ Mpc\cite{2018Natur.554..497B}.

\subsection{A Hybrid shock cooling + interaction model}\label{sec:hybrid}
Approximate analytic expressions for the optical light curves following shock breakout of the stellar envelope have been investigated by various studies \cite{2010ApJ...725..904N, 2011ApJ...728...63R, 2013ApJ...769...67P, 2014ApJ...788..193N, 2017ApJ...838..130S}. 
Following the shock breakout from an RSG progenitor, the early shock-cooling emission can be characterized by two stages: the `planar' and the `spherical' phases. The `planar' phase lasts for about a few hours, contributing to the radiation soon after the explosion, which then transitions to the `spherical' phase with a duration of several days.

\subsubsection{Shock cooling} ~\label{sec:shock_cooling}
Within the initial dynamical time of the breakout shell, only a small fraction of emission leaks out the envelope. This initial dynamical time is about a few minutes, and is earlier than the date of our first photometric data according to the first-light time fitted by either the broken-power model (Fig.~\ref{fig:fireball}) or shock cooling process (Fig.~\ref{fig:modelfit_contour}). Therefore, only the subsequent stages ('planar' and 'spherical phases') of the shock cooling are considered in our model as below:

1) Immediately after the generation of the shock wave, the outermost stellar layers will remain rather static until being caught by the breakout front. For a progenitor with an initial radius of $R_{\rm star}$, when $t < R_{\rm star} / v_{\rm shock,*}$, the photons will only be able to escape within a thin, surface shell with an optical depth of $\tau \sim c / v_{\rm shock,*}$, where $v_{\rm shock,*}$ describes the shock velocity as it breaks out of the stellar surface. 
Therefore, $\tau$ remains constant, and the thickness of the emitting shell $\Delta$$r$ extends proportionally with $t$ \cite{2010ApJ...708..598P, 2010ApJ...725..904N}. Such an expansion can be approximated in the plane-parallel limit before the shock energy injects into the surface envelope. The resultant luminosity and temperature evolution can be written as\cite{2010ApJ...725..904N} $L(t) \propto t_{\rm day}^{-4/3}$ and $T(t) \propto t_{\rm day}^{-0.36}$, where `$t_{\rm day}$' represents the post-explosion time in unit of a day; \\
2) When the shock propagates through the outer envelope of the star, i.e., $t > R_{\rm star} / v_{\rm shock,*}$, the gas envelopes start to expand homologously, with different expansion velocities at different layers. 
During this subsequent `spherical phase', the observed luminosity yields\cite{2010ApJ...725..904N}: 

\begin{align}
L = 3\times10^{42} {\rm erg}\ {\rm s}^{-1} M^{-0.87}_{\rm 15} R_{500} E^{0.96}_{51} t_{\rm day}^{-0.17}, \\
T = 3\ {\rm eV} M_{15}^{-0.13} R_{500}^{0.38} E_{51}^{0.11} t_{\rm day}^{-0.56},
\end{align}

where $M_{15}$ and $R_{500}$ represent the mass and the initial radius of the exploding RSG in units of 15$M_{\odot}$ and 500$R_{\odot}$, respectively. The explosion energy, $E_{51}$, is denoted in unit of 10$^{51}$ erg. For a common RSG with the above parameter units, the planar-to-spherical phase transition takes place at\cite{2010ApJ...725..904N} 

\begin{align}
t_{s} = 14\ {\rm hr}\, M_{15}^{0.43} R_{500}^{1.26} E_{51}^{-0.56}.
\end{align}

Towards the end of the shock cooling phase, the luminosity will be affected by the treatment of the emission envelope as the emission from deeper layers becomes progressively significant. The depression of the observed flux can be explained by our modeling with the analytic approximation given by \cite{2017ApJ...838..130S}.

\subsubsection{Ejecta-CSM interaction}~\label{sec:interaction} 
The immediate SN environment is determined by the wind velocity ($v_{\text{w}}$) and the mass-loss rate ($\dot{M}_{\text{w}}$), making a radial density profile of $\rho_{\rm csm} = \dot{M}_{\text{w}} / 4 \pi R^{2} v_{\text{w}}$, where $R$ is the distance from the SN. The dynamic procedure of the ejecta smashing into the CSM can be solved as\cite{2003LNP...598..171C}: 

\begin{equation} 
M_{\text{sh}}\frac{\mathrm{d}V_{\text{sh}}}{\mathrm{d}t} = 4\pi R_{\text{sh}}^2[\rho_{\text{ej}}(v_{\text{ej}} - V_{\text{sh}})^2 - \rho_{\text{csm}}(V_{\text{sh}} - v_{\text{w}})^2], 
\end{equation} 

where $M_{\rm sh}$, $V_{\rm sh}$, and $R_{\rm sh}$ give the mass, expanding velocity, and radius of the layer where the ejecta collides with the CSM, respectively. 
The ejecta expands at a velocity $v_{\rm ej}$ and follows a power-law radial density distribution of $\rho_{\rm ej} \propto R^{-n}$. The interaction luminosity thus gives 

\begin{equation}
L = \frac{\epsilon}{2}\frac{\dot{M}_{\text{w}}}{v_{\rm w}}V_{\text{sh}}^3, 
 \end{equation}

where $\epsilon$ is the conversion efficiency from kinetic energy to radiation, which has a typical value of 0.15.

To examine the time-variant mass ejection during the final intense mass-loss phase, we adopted the following mass-loss profile that manifests differently across the CSM radial density profile \cite{Hu2023_CSMSNeIa}:

\begin{align} 
\label{eq_Mw}
\dot{M}_{\text{w}}(R) = \begin{cases} 
0, & R \le R_0 \\
\dot{M}_{\text{w}}(0)(\frac{R - R_0}{R_1 - R_0})^{n_1}, & R_0 < R \le R_1 \\ 
\dot{M}_{\text{w}}(0), & R_1 < R \le R_2 \\
\dot{M}_{\text{w}}(0) (\frac{R_3 - R}{R_3 - R_2})^{n_2}, & R_2 < R \le R_3 \\
\end{cases}
\end{align} 
The CSM profile shown in Eq.~\ref{eq_Mw} yields no CSM within the radius of $R_{0}$ from the SN, leading to a delay time of $\sim R_{0} / V_{\mathrm{sh}}$ for the ejecta$-$CSM interaction. The trend of increasing-constant-decreasing mass-loss rate $\dot{M_{\rm w}}(R)$ from $R_{0}$ to $R_{3}$ characterizes an eruption episode, corresponding to
an intense mass-loss that starts ($R_{0,3} = R_{3}$) and terminates ($R_{0,3} = R_{0}$) at $\frac{R_{0,3} - R_{\rm star}}{v_{\rm w}} \approx 116\,{\rm day} (\frac{R_{0,3}}{10^{14}\, \rm cm} - \frac{R_{\rm star}}{450\,R_{\odot}}\cdot 0.313)/(\frac{v_{\rm w}}{100\,\rm km \, s^{-1}})$ before the final explosion.

\subsubsection{Time-variant extinction and dust grain sublimation}~\label{sec:extinction}
Analysis of the optical-to-mid infrared data
suggests the presence of a dusty shell around the RSG progenitor of SN\,2023ixf\cite{2023ApJ...952L..23K}. Any dust grains of proximity could be quickly accelerated and sublimated by the energetic X-ray/UV radiation ($\sim10^{43}$ erg s$^{-1}$) followed by the shock breakout.

Depending on the shock velocity and properties of the dust shell, at a given time after the SN explosion, dust grains located beyond a distance of $\sim v_{\rm shock}t$ may survive early on and remarkably modify the luminosity and color within the first hours to days. With the presence of CSM, the arrival time of photons varies with their emergent positions at the `effective' photosphere, as defined by $\tau \sim c/v_{\rm shock}$. The LOS photons arrive first, and the locus of constant photon-travel time is defined by an iso-$\tau$ surface that intersects the aforementioned photosphere. 
For a qualitative analysis of the aftermath of an aspherical shock breakout, we hereby estimate the actual luminosity received by a dust shell at $d_{0}$ from the emission center of the SN.

At a given time $t$, the iso-delay emission surface intersects the shock breakout surface at a ring that has a distance $R$ to the SN emission center and a polar angle $\theta$. Considering $\Delta t$ to be the time delay of the emission from the ring ($R$, $\theta$) compared to that from the LOS (i.e., $\theta=0$), we can write $R = R(t-\Delta t, \theta)$. 
For a dust clump located above the breakout surface and at a distance of $d_{0}$ along the SN-Earth direction, given a shock-breakout intensity $I(\theta, t)$, the corresponding flux gives:

\begin{equation}~\label{eqn:dust}
F(t, d_{0}) = 2\int 
\frac{I(t-\Delta t, \theta) \cdot R(t -\Delta t, \theta)^{2}}{d(t - \Delta t, \theta)^2} 
\cos(\alpha)\,\mathrm{d}(\cos(\theta)), 
\end{equation}
where $\alpha$ is the angle between the radial direction on the ring of $R(t-\Delta t, \theta)$ and the position of the dust cloud seen from a given point of the ring, and $d(t - \Delta t, \theta)$ is the distance between the ring and dust clump. For the sake of simplicity, we adopt the following assumptions: \\
1) The shock breakout follows an oblate geometry. As a consequence, the breakout front moving towards the observer is the slowest compared to other directions, thus causing a delay in heating the LOS dust grains and hence allowing them to survive for a longer period; \\
2) The emission from an aspherical shock breakout follows a thermal spectrum from adiabatically-cooling material within the shell of the breakout. 

Therefore, the dust extinction will exist at $t_{0}$ until most of the grains are sublimated. Adopting the optical properties calculated for graphite grains with radius $a=0.05\,\mu$m \cite{1984ApJ...285...89D, 1993ApJ...402..441L, 2001ApJ...548..296W,2022ApJ...931..110H}, we follow the prescription of \cite{1988ApJ...329..814D} to compute the survival time of grains at different temperatures. With the heating of such off-LOS aspherical breakout front, the LOS grains at a distance of $\sim5\times10^{14}$ to $5\times10^{15}$ cm from the SN may survive a few minutes to $\sim$5 hours for a typical oblate breakout front with an axis ratio of $\approx$10.

Despite the approximate duration of dust survival time agrees with the short-lived extinction term required to account for flux suppression seen in the blue band within the first five hours (see  Fig.~\ref{fig:modelfit}), a more careful treatment of the emission from shock-cooling due to an aspherical shock breakout, together with detailed geometry of the breakout front, are essential to characterize the effect caused by such a rapidly time-varying extinction. 

\subsection{\textbf{Hints to an Aspherical Shock Breakout}}~\label{sec:aspherical}
An aspherical breakout inside an extended, optically thick CSM will produce the first light from the direction with the least amount of CSM. The CSM dust may survive and manifest as extinctions along directions with larger optical depth. As the breakout emerges from more directions, the overlying dust gets gradually sublimated and the integrated extinction tends to decrease over time.

The aspherical shock breakout may wrap around the stellar surface, starting from  where the breakout acceleration takes place \cite{2021MNRAS.508.5766I}.  
Depending on the inclination angle of the shock relative to the stellar surface, the degree of asphericity, and the location of the initial breakout, the pattern velocity $v_p$ \cite{2021MNRAS.508.5766I}, which describes how fast the shock propagates, may significantly modify the early-time emission. For example, a more prolonged and weaker shock breakout will be produced with an increased asphericity \cite{2018ApJ...856..146A, 2021MNRAS.508.5766I,2013ApJ...779...60M}. In the highly aspherical regime, the peak luminosity can be up to $\sim$2 orders of magnitude fainter than that determined for an adiabatic shock-cooling with spherical symmetry, and the duration of such a breakout can be extended up to $\sim$10 hours for an RSG star. An aspherical and optically thick CSM around SN\,2023ixf has already been suggested by the early spectroscopic evolution \cite{2023ApJ...956...46S} and spectropolarimetry \cite{2023ApJ...955L..37V}.

The wrapping-around procedure would result in a complex interplay between multiple time-varying emission sources and sublimation of dust grains within $\sim$hr timescales. 
Assuming an oblate shock breakout that leads to delayed heating of the LOS CSM, we estimate that dust grains may survive the first few hours at a distance of a few $\sim10^{14}$ to $10^{15}$ cm from the SN (see Method~\ref{sec:extinction}). 
Very recent 3D hydrodynamic simulations also suggest a prolonged and fainter breakout that exhibits remarkable large-scale anisotropy\cite{2022ApJ...933..164G}. 
For a typical RSG with a few hundred R$_{\odot}$, the breakout duration extends to $\approx 3-6$ hrs while the luminosity decreases by a factor of 3--10 times, evincing an increased diffusion time due to low-density material occupying a substantial amount of space above the photosphere, and the presence of large-scale (i.e., $\sim100\,R_{\odot}$) plumes of density fluctuations. These implications of 3D calculations would suggest more relaxed conditions for the survival of any ambient attenuating dust grains.

These claims are compatible with the evolution of the extinction term that is required to account for the early suppression of the blue emission within hours after the first light, further strengthening an aspherical shock breakout.
Additional dust destruction mechanisms such as rains spun-up and disruptions by strong radiative torques\cite{2019NatAs...3..766H}) may allow a $\sim$1--2 hr survival time for any dust grains with a radius of $\sim0.1\,\mu$m within $5\times 10^{14}$\,cm of SN\,2023ixf.

\clearpage
\section{Data availability}
Raw images of SN\,2023ixf obtained by the 13-cm Newtonian, 10.1-cm refractor, and 12-inch RC telescope observed by co-authors can be retrieved from the Transient Name Server (\url{https://www.wis-tns.org/}). The images of the 0.6-m NEXT in Xingjiang, China, the 10.7-cm refractor in Xinjiang, China, the 15.2-cm refractor in Sola, Norway, the 13-cm refractor in Yunnan, China, the 10-cm refractor in Yunnan, China and the 15.0-cm refractor in Utah, USA, are exclusive or published for the first time. All photometric data except those from LT, PT, and ZTF are original. All reduced light curves used for this work are available on our Zenodo page (  \url{https://doi.org/10.5281/zenodo.8434500}).

\section{Code availability}
The {\sc python} package \textsc{reproject} (v1.2.0) for stacking images can be obtained from \url{https://reproject.readthedocs.io/en/stable/index.html}.
The \textsc{PyZOGY} package and \textsc{SFFT} package used for image subtraction are available at \url{https://github.com/dguevel/PyZOGY} and \url{https://github.com/thomasvrussell/sfft} respectively.
The codes of {\sc AUTOPHOT} that are used for reducing photometric images are available at \url{https://github.com/Astro-Sean/autophot}.
The program used for aperture photometry to reduce AST3-3 and YAHPT images is SourceExtractor, which can be downloaded from \url{https://www.astromatic.net/software/sextractor/}.
The general tool used in this study to solve the  World Coordinate System (WCS) is astrometry.net \url{http://astrometry.net/doc/build.html#build}. The code to fit the hybrid model is available on \url{https://doi.org/10.5281/zenodo.8434500}

\section{Acknowledgments}
The work of X.-F.W. is supported by the National Science Foundation of China (NSFC grants 12288102, 12033003, and 11633002), the Scholar Program of Beijing Academy of Science and Technology (DZ: BS202002), and New Cornerstone Science Foundation through the XPLORER PRIZE. 
M.-K.H. acknowledges support from the National Natural Science Foundation of China (grant Nos. 12321003) and the Jiangsu Funding Program for Excellent Postdoctoral Talent. 
W.-X.L. acknowledges support from an Israel Science Foundation (ISF) grant (number 2752/19), a European Research Council (ERC) grant (JetNS) under the European Union's Horizon 2020 research and innovation program, and the National Natural Science Foundation of China (NSFC; grant Nos. 12120101003 and 12233008).
Y.Y. appreciates the generous financial support provided to the supernova group at U.C. Berkeley by Gary and Cynthia Bengier, Clark and Sharon Winslow, Sanford Robertson, and numerous other donors.
L.H. acknowledges support from Jiangsu Funding Program for Excellent Postdoctoral Talent, the~Major Science and Technology Project of Qinghai Province (2019-ZJ-A10) and China Postdoctoral Science Foundation (Grant No. 2022M723372).
T.-R.S. appreciate the Major Science and Technology Project of Qinghai Province (2019-ZJ-A10), and the Jiangsu Funding Program for Excellent Postdoctoral Talent.
J.-J.Z. is supported by the National Key R\&D Program of China with No. 2021YFA1600404, the National Natural Science Foundation of China (12173082), the Yunnan Province Foundation (202201AT070069), the Top-notch Young Talents Program of Yunnan Province, the Light of West China Program provided by the Chinese Academy of Sciences, the International Centre of Supernovae, Yunnan Key Laboratory (No. 202302AN360001).
D.-R.X. acknowledges the support of BOOTES-4 technical staffs. L.-Z.W. is sponsored (in part) by the Chinese Academy of Sciences (CAS), through a grant to the CAS South America Center for Astronomy (CASSACA) in Santiago, Chile.
Jian Chen acknowledges support from the National Natural Science Foundation of China (Grant Nos. 12203105).

The AST3-3 team and YAHPT team would like to express their sincere thanks to the staff of the Yaoan observation station.
The operation of Xingming Observatory is supported by the Education Bureau of Ningbo, China and Xinjiang Astronomical Observatory, China.
We acknowledge the support of the staff of the Xinglong
80cm telescope. This work was partially supported by the Open Project
Program of the CAS Key Laboratory of Optical Astronomy, National Astronomical
Observatories, Chinese Academy of Sciences.
The SNOVA team is supported by the National Key R\&D program of China for Intergovernmental Scientific and Technological Innovation Cooperation Project under No. 2022YFE0126200, High-Level Talent-Heaven Lake Program of Xinjiang Uygur Autonomous Region of China.

\section*{Author contributions}
X.-F. W., Y. Y., G.-C. L., M.-K. H., and W.-X. L. drafted the manuscript. X.-F. W. initiated this study and led the discussions. G.-C.L. and W.-X. L. led the data reduction and analysis. M.-K. H. and Y. Y. led the hybrid model fitting. S.-Y.Y. contributed to the fitting of the early-time light curves. L.H. helped with the AST3 data reduction and the manuscript; T.-R.S. helped with the YAHPT data reduction; D.-R. X. and J.-J.Z. helped with the BOOTES data reduction; J.-L.L. and J. M. helped with the TNT data reduction; A. I. helped with the SNOVA data reduction. 
Y.-M.M. obtained the data with the 10.7-cm refractor/ZWO ASI071MC-Cool camera; H.R. obtained the data with the 15.2-cm refractor/ZWO ASI6200MM Pro camera; V.C., N.P., I.I., S.K., S.N., and K.S. obtained the data with the 10.1-cm refractor/QHY600m monochrome CMOS camera; X.G. obtained the data with the 0.6-m reflector/FLI 230-42 camera; Jian Chen obtained the data with YAHPT; T.-R.S. obtained the data with AST3; G.-Y.C. obtained the data with the 13-cm refractor/ZWO ASI2600MM Pro; Jin Chen obtained the data with the 10-cm refractor/ZWO ASI2600MM; N.H. obtained the data with the 12-inch Ritchey–Chr\'etien (RC) telescope/ATR3CMOS 26000KPA; E.H. obtained the data with the 15-cm refractor/FLI Proline 10002M. 
G.-B.X., D.-F.X., L-F.W., L.-Z.W., W.-L.L., F.-Z.G.,and Q.-C.L. helped with the discussions. K.-M.Z., G.-Y.S., W.-J. Z., J.-Y.Z., X.Z., K.-Y.L., M.Z., S.-J.X., X.-D.L. and A.J.C.-T. contributed to the data collections. 

\section*{Competing interests}
The authors declare no competing interests.

\clearpage
\begin{longtable}{ccccc}
  \caption{Photometric Observations of SN~2023ixf before MJD 60100.}\label{tab:photometry}\\
\hline
MJD & Phase$^a$ & Filter & Magnitude & Instrument/Contributor\\
\hline
\endfirsthead
\multicolumn{5}{c}%
{{\bfseries \tablename\ \thetable{} -- Continued}} \\
\hline 
MJD & Phase$^a$ & Filter & Magnitude & Instrument\\
\hline
\endhead
\hline 
\multicolumn{5}{r}{{Continued}} \\ \hline
\endfoot
\hline
\hline
\endlastfoot
60081.753	&	-1.035	&	$r$	&	$>20.5$	&		NEXT\\
60082.818	&	0.030	&	$r$	&	$>18.6$	&		13-cm	Newtonian\\	
60082.829	&	0.041	&	$r$	&	18.101$\pm$0.343 	&	13-cm	Newtonian	\\	
60082.847	&	0.059	&	$r$	&	17.838$\pm$0.312 	&	10.7-cm	refractor	\\	
60082.876	&	0.088	&	$r$	&	17.685$\pm$0.164 	&	10.7-cm	refractor	\\	
60082.882	&	0.094	&	$r$	&	17.708$\pm$0.320 	&	13-cm	Newtonian	\\	
60082.899	&	0.111	&	$r$	&	17.623$\pm$0.270 	&	13-cm	Newtonian	\\	
60082.906	&	0.118	&	$r$	&	17.650$\pm$0.225 	&	12-inch	RC	telescope	\\
60082.941	&	0.153	&	$r$	&	17.461$\pm$0.119 	&	10.1-cm	refractor	\\	
60082.944	&	0.156	&	$r$	&	17.450$\pm$0.139 	&	12-inch	RC	telescope	\\	
60082.947	&	0.159	&	$r$	&	17.468$\pm$0.123 	&	15.2-cm	refractor	\\	
60083.024	&	0.236	&	$r$	&	17.079$\pm$0.063 	&	12-inch	RC	telescope	\\	
60084.312	&	1.524	&	$r$	&	13.659$\pm$0.100	&	15-cm refractor	\\		
60084.613	&	1.825	&	$r$	&	12.836$\pm$0.021	&	BOOTES-4	\\		
60084.626	&	1.838	&	$r$	&	12.793$\pm$0.021	&	BOOTES-4	\\		
60084.630	&	1.842	&	$r$	&	12.848$\pm$0.021	&	BOOTES-4	\\		
60084.634	&	1.846	&	$r$	&	12.774$\pm$0.021	&	BOOTES-4	\\		
60084.637	&	1.849	&	$r$	&	12.841$\pm$0.021	&	BOOTES-4	\\		
60084.640	&	1.852	&	$r$	&	12.828$\pm$0.021	&	BOOTES-4	\\		
60084.645	&	1.857	&	$r$	&	12.764$\pm$0.021	&	BOOTES-4	\\		
60084.697	&	1.909	&	$r$	&	12.668$\pm$0.021	&	BOOTES-4	\\		
60084.699	&	1.911	&	$r$	&	12.709$\pm$0.021	&	BOOTES-4	\\		
60084.701	&	1.913	&	$r$	&	12.683$\pm$0.021	&	BOOTES-4	\\		
60084.703	&	1.915	&	$r$	&	12.678$\pm$0.021	&	BOOTES-4	\\		
60084.705	&	1.917	&	$r$	&	12.703$\pm$0.021	&	BOOTES-4	\\		
60084.752	&	1.964	&	$r$	&	12.609$\pm$0.021	&	BOOTES-4	\\		
60084.755	&	1.967	&	$r$	&	12.620$\pm$0.021	&	BOOTES-4	\\		
60084.758	&	1.970	&	$r$	&	12.598$\pm$0.021	&	BOOTES-4	\\		
60084.760	&	1.972	&	$r$	&	12.598$\pm$0.021	&	BOOTES-4	\\		
60084.764	&	1.976	&	$r$	&	12.581$\pm$0.021	&	BOOTES-4	\\		
60084.904	&	2.116	&	$r$	&	12.459$\pm$0.021	&	BOOTES-4	\\		
60085.586	&	2.798	&	$r$	&	11.765$\pm$0.016	&	BOOTES-4	\\		
60085.590	&	2.802	&	$r$	&	11.762$\pm$0.016	&	BOOTES-4	\\		
60085.595	&	2.807	&	$r$	&	11.753$\pm$0.016	&	BOOTES-4	\\		
60085.599	&	2.811	&	$r$	&	11.737$\pm$0.016	&	BOOTES-4	\\		
60085.604	&	2.816	&	$r$	&	11.753$\pm$0.016	&	BOOTES-4	\\		
60085.609	&	2.821	&	$r$	&	11.755$\pm$0.016	&	BOOTES-4	\\		
60085.615	&	2.827	&	$r$	&	11.742$\pm$0.016	&	BOOTES-4	\\		
60085.622	&	2.834	&	$r$	&	11.757$\pm$0.016	&	BOOTES-4	\\		
60085.628	&	2.840	&	$r$	&	11.745$\pm$0.016	&	BOOTES-4	\\		
60085.698	&	2.910	&	$r$	&	11.691$\pm$0.016	&	BOOTES-4	\\		
60085.702	&	2.914	&	$r$	&	11.687$\pm$0.016	&	BOOTES-4	\\		
60085.705	&	2.917	&	$r$	&	11.684$\pm$0.016	&	BOOTES-4	\\		
60085.708	&	2.920	&	$r$	&	11.701$\pm$0.016	&	BOOTES-4	\\		
60086.590	&	3.802	&	$r$	&	11.389$\pm$0.021	&	BOOTES-4	\\		
60086.604	&	3.816	&	$r$	&	11.374$\pm$0.021	&	BOOTES-4	\\		
60086.609	&	3.821	&	$r$	&	11.370$\pm$0.021	&	BOOTES-4	\\		
60086.616	&	3.828	&	$r$	&	11.361$\pm$0.021	&	BOOTES-4	\\		
60086.621	&	3.833	&	$r$	&	11.370$\pm$0.021	&	BOOTES-4	\\		
60086.689	&	3.901	&	$r$	&	11.345$\pm$0.021	&	BOOTES-4	\\		
60086.691	&	3.903	&	$r$	&	11.345$\pm$0.021	&	BOOTES-4	\\		
60086.694	&	3.906	&	$r$	&	11.335$\pm$0.021	&	BOOTES-4	\\		
60086.696	&	3.908	&	$r$	&	11.355$\pm$0.021	&	BOOTES-4	\\		
60086.699	&	3.911	&	$r$	&	11.345$\pm$0.021	&	BOOTES-4	\\		
60087.564	&	4.776	&	$r$	&	11.245$\pm$0.012	&	BOOTES-4	\\		
60087.570	&	4.782	&	$r$	&	11.245$\pm$0.012	&	BOOTES-4	\\		
60087.579	&	4.791	&	$r$	&	11.245$\pm$0.012	&	BOOTES-4	\\		
60088.551	&	5.763	&	$r$	&	11.195$\pm$0.010	&	BOOTES-4	\\		
60088.579	&	5.791	&	$r$	&	11.195$\pm$0.010	&	BOOTES-4	\\		
60088.583	&	5.795	&	$r$	&	11.197$\pm$0.010	&	BOOTES-4	\\		
60088.586	&	5.798	&	$r$	&	11.174$\pm$0.010	&	BOOTES-4	\\	
60089.673	&	6.885	&	$r$	&	11.221$\pm$0.038	&	TNT	\\		
60096.570	&	13.782	&	$r$	&	11.187$\pm$0.031	&	BOOTES-4	\\		
60096.576	&	13.788	&	$r$	&	11.171$\pm$0.031	&	BOOTES-4	\\		
60096.581	&	13.793	&	$r$	&	11.178$\pm$0.031	&	BOOTES-4	\\		
60096.586	&	13.798	&	$r$	&	11.191$\pm$0.031	&	BOOTES-4	\\		
60097.576	&	14.788	&	$r$	&	11.194$\pm$0.015	&	BOOTES-4	\\		
60097.594	&	14.806	&	$r$	&	11.197$\pm$0.015	&	BOOTES-4	\\		
60097.598	&	14.810	&	$r$	&	11.211$\pm$0.015	&	BOOTES-4	\\		
60097.603	&	14.815	&	$r$	&	11.201$\pm$0.015	&	BOOTES-4	\\		
60097.608	&	14.820	&	$r$	&	11.198$\pm$0.015	&	BOOTES-4	\\		
60098.572	&	15.784	&	$r$	&	11.201$\pm$0.028	&	BOOTES-4	\\		
60098.577	&	15.789	&	$r$	&	11.227$\pm$0.028	&	BOOTES-4	\\		
60098.582	&	15.794	&	$r$	&	11.210$\pm$0.028	&	BOOTES-4	\\		
60098.587	&	15.799	&	$r$	&	11.216$\pm$0.028	&	BOOTES-4	\\		
60098.592	&	15.804	&	$r$	&	11.207$\pm$0.028	&	BOOTES-4	\\		
60099.550	&	16.762	&	$r$	&	11.245$\pm$0.031	&	BOOTES-4	\\		
60099.556	&	16.768	&	$r$	&	11.241$\pm$0.031	&	BOOTES-4	\\		
60099.561	&	16.773	&	$r$	&	11.228$\pm$0.031	&	BOOTES-4	\\		
60099.567	&	16.779	&	$r$	&	11.239$\pm$0.031	&	BOOTES-4	\\		
60099.572	&	16.784	&	$r$	&	11.261$\pm$0.031	&	BOOTES-4	\\		
60100.686	&	17.898	&	$r$	&	11.258$\pm$0.024	&	TNT	\\		
60082.847	&	0.059	&	$V$	&	18.360$\pm$0.281 	&	10.7-cm	refractor\\		
60082.876	&	0.088	&	$V$	&	18.141$\pm$0.215 	&	10.7-cm	refractor	\\	
60082.889	&	0.101	&	$V$	&	18.041$\pm$0.110 	&	12-inch	RC	telescope	\\
60082.906	&	0.118	&	$V$	&	17.802$\pm$0.115 	&	12-inch	RC	telescope	\\
60082.942	&	0.154	&	$V$	&	17.652$\pm$0.112 	&	10.1-cm	refractor	\\	
60082.944	&	0.156	&	$V$	&	17.678$\pm$0.075 	&	12-inch	RC	telescope	\\
60082.974	&	0.186	&	$V$	&	17.460$\pm$0.156 	&	15.2-cm	refractor	\\	
60083.024	&	0.236	&	$V$	&	17.148$\pm$0.059 	&	12-inch	RC	telescope	\\
60083.585	&	0.797	&	$V$	&	15.065$\pm$0.048 	&	10-cm refractor	\\
60083.592	&	0.804	&	$V$	&	15.028$\pm$0.076 	&	10-cm refractor	\\
60083.614	&	0.826	&	$V$	&	14.984$\pm$0.080 	&	10-cm refractor	\\
60083.617	&	0.829	&	$V$	&	15.061$\pm$0.029 	&	13-cm	refractor	\\	
60083.617	&	0.829	&	$V$	&	14.919$\pm$0.076 	&	10-cm refractor	\\
60084.312	&	1.524	&	$V$	&	13.399$\pm$0.063	&	15-cm refractor	\\		
60084.538	&	1.750	&	$V$	&	13.10$\pm$0.01	&	YAHPT	\\		
60084.539	&	1.751	&	$V$	&	13.09$\pm$0.01	&	YAHPT	\\		
60084.540	&	1.752	&	$V$	&	13.09$\pm$0.01	&	YAHPT	\\		
60084.576	&	1.788	&	$V$	&	13.02$\pm$0.01	&	YAHPT	\\		
60084.599	&	1.811	&	$V$	&	12.98$\pm$0.01	&	YAHPT	\\		
60084.651	&	1.863	&	$V$	&	12.87$\pm$0.01	&	YAHPT	\\		
60085.616	&	2.828	&	$V$	&	11.86$\pm$0.01	&	YAHPT	\\		
60085.700	&	2.912	&	$V$	&	11.78$\pm$0.01	&	YAHPT	\\		
60089.544	&	6.756	&	$V$	&	11.32$\pm$0.01	&	YAHPT	\\		
60089.606	&	6.818	&	$V$	&	11.32$\pm$0.01	&	YAHPT	\\		
60091.637	&	8.849	&	$V$	&	11.35$\pm$0.01	&	YAHPT	\\		
60091.728	&	8.940	&	$V$	&	11.33$\pm$0.01	&	YAHPT	\\		
60091.822	&	9.034	&	$V$	&	11.34$\pm$0.01	&	YAHPT	\\		
60092.750	&	9.962	&	$V$	&	11.34$\pm$0.01	&	YAHPT	\\		
60094.650	&	11.862	&	$V$	&	11.36$\pm$0.01	&	YAHPT	\\		
60094.743	&	11.955	&	$V$	&	11.36$\pm$0.01	&	YAHPT	\\		
60095.702	&	12.914	&	$V$	&	11.37$\pm$0.01	&	YAHPT	\\		
60096.650	&	13.862	&	$V$	&	11.40$\pm$0.01	&	YAHPT	\\		
60096.746	&	13.958	&	$V$	&	11.41$\pm$0.01	&	YAHPT	\\		
60097.648	&	14.860	&	$V$	&	11.42$\pm$0.01	&	YAHPT	\\		
60097.754	&	14.966	&	$V$	&	11.42$\pm$0.01	&	YAHPT	\\
60098.689	&	15.901	&	$V$	&	11.448$\pm$0.002	&	SNOVA	\\		
60098.795	&	16.007	&	$V$	&	11.447$\pm$0.003	&	SNOVA	\\
60099.549	&	16.761	&	$V$	&	11.49$\pm$0.01	&	YAHPT	\\		
60099.806	&	17.018	&	$V$	&	11.51$\pm$0.01	&	YAHPT	\\		
60100.584	&	17.796	&	$V$	&	11.51$\pm$0.01	&	YAHPT	\\		
60082.847	&	0.059	&	$g$	&	$>18.5$ 		&	10.7-cm	refractor	\\	
60082.876	&	0.088	&	$g$	&	$>18.5$ 		&	10.7-cm	refractor	\\	
60082.889	&	0.101	&	$g$	&	18.214$\pm$0.147	&	12-inch	RC	telescope	\\
60082.906	&	0.118	&	$g$	&	18.000$\pm$0.115	&	12-inch	RC	telescope	\\
60082.944	&	0.156	&	$g$	&	17.697$\pm$0.108 		&	12-inch	RC	telescope	\\
60082.945	&	0.157	&	$g$	&	17.534$\pm$0.113 	&	10.1-cm	refractor	\\	
60082.964	&	0.176	&	$g$	&	17.441$\pm$0.138 		&	15.2-cm	refractor	\\	
60083.024	&	0.236	&	$g$	&	17.130$\pm$0.082 		&	12-inch	RC	telescope	\\
60084.312	&	1.524	&	$g$	&	13.107$\pm$0.098	&	15-cm refractor	\\		
60084.557	&	1.769	&	$g$	&	12.643$\pm$0.012	&	AST3-3	\\		
60084.558	&	1.770	&	$g$	&	12.637$\pm$0.012	&	AST3-3	\\		
60084.598	&	1.810	&	$g$	&	12.563$\pm$0.012	&	AST3-3	\\		
60084.599	&	1.811	&	$g$	&	12.558$\pm$0.012	&	AST3-3	\\		
60084.640	&	1.852	&	$g$	&	12.489$\pm$0.012	&	AST3-3	\\		
60084.641	&	1.853	&	$g$	&	12.479$\pm$0.012	&	AST3-3	\\		
60084.682	&	1.894	&	$g$	&	12.418$\pm$0.012	&	AST3-3	\\		
60084.683	&	1.895	&	$g$	&	12.400$\pm$0.012	&	AST3-3	\\		
60084.723	&	1.935	&	$g$	&	12.341$\pm$0.012	&	AST3-3	\\		
60084.724	&	1.936	&	$g$	&	12.337$\pm$0.012	&	AST3-3	\\		
60085.558	&	2.770	&	$g$	&	11.475$\pm$0.011	&	AST3-3	\\		
60085.558	&	2.770	&	$g$	&	11.473$\pm$0.011	&	AST3-3	\\		
60085.599	&	2.811	&	$g$	&	11.442$\pm$0.011	&	AST3-3	\\		
60085.599	&	2.811	&	$g$	&	11.437$\pm$0.011	&	AST3-3	\\		
60085.641	&	2.853	&	$g$	&	11.427$\pm$0.011	&	AST3-3	\\		
60085.641	&	2.853	&	$g$	&	11.419$\pm$0.011	&	AST3-3	\\		
60085.682	&	2.894	&	$g$	&	11.402$\pm$0.011	&	AST3-3	\\		
60085.683	&	2.895	&	$g$	&	11.395$\pm$0.011	&	AST3-3	\\		
60085.724	&	2.936	&	$g$	&	11.380$\pm$0.011	&	AST3-3	\\		
60085.724	&	2.936	&	$g$	&	11.380$\pm$0.011	&	AST3-3	\\		
60086.557	&	3.769	&	$g$	&	11.120$\pm$0.012	&	AST3-3	\\		
60086.557	&	3.769	&	$g$	&	11.104$\pm$0.012	&	AST3-3	\\		
60086.598	&	3.810	&	$g$	&	11.094$\pm$0.011	&	AST3-3	\\		
60086.598	&	3.810	&	$g$	&	11.093$\pm$0.011	&	AST3-3	\\		
60086.681	&	3.893	&	$g$	&	11.073$\pm$0.011	&	AST3-3	\\		
60086.682	&	3.894	&	$g$	&	11.075$\pm$0.011	&	AST3-3	\\		
60087.551	&	4.763	&	$g$	&	10.969$\pm$0.011	&	AST3-3	\\		
60087.608	&	4.820	&	$g$	&	10.982$\pm$0.011	&	AST3-3	\\		
60087.630	&	4.842	&	$g$	&	10.972$\pm$0.011	&	AST3-3	\\
60088.549	&	5.761	&	$g$	&	10.890$\pm$0.038	&	BOOTES-4	\\		
60088.581	&	5.793	&	$g$	&	10.924$\pm$0.038	&	BOOTES-4	\\
60088.585	&	5.797	&	$g$	&	10.900$\pm$0.038	&	BOOTES-4	\\		
60089.607	&	6.819	&	$g$	&	10.984$\pm$0.011	&	AST3-3	\\		
60089.658	&	6.870	&	$g$	&	10.987$\pm$0.011	&	AST3-3	\\		
60090.580	&	7.792	&	$g$	&	11.027$\pm$0.011	&	AST3-3	\\		
60090.664	&	7.876	&	$g$	&	11.027$\pm$0.011	&	AST3-3	\\		
60091.632	&	8.844	&	$g$	&	11.052$\pm$0.011	&	AST3-3	\\		
60091.642	&	8.854	&	$g$	&	11.075$\pm$0.011	&	AST3-3	\\		
60091.685	&	8.897	&	$g$	&	11.071$\pm$0.011	&	AST3-3	\\		
60091.697	&	8.909	&	$g$	&	11.048$\pm$0.011	&	AST3-3	\\		
60093.572	&	10.784	&	$g$	&	11.092$\pm$0.011	&	AST3-3	\\		
60093.623	&	10.835	&	$g$	&	11.090$\pm$0.011	&	AST3-3	\\		
60093.650	&	10.862	&	$g$	&	11.078$\pm$0.011	&	AST3-3	\\		
60093.662	&	10.874	&	$g$	&	11.090$\pm$0.011	&	AST3-3	\\		
60093.686	&	10.898	&	$g$	&	11.088$\pm$0.011	&	AST3-3	\\		
60094.621	&	11.833	&	$g$	&	11.092$\pm$0.011	&	AST3-3	\\		
60094.648	&	11.860	&	$g$	&	11.111$\pm$0.011	&	AST3-3	\\		
60094.675	&	11.887	&	$g$	&	11.129$\pm$0.011	&	AST3-3	\\		
60094.706	&	11.918	&	$g$	&	11.133$\pm$0.011	&	AST3-3	\\		
60095.574	&	12.786	&	$g$	&	11.136$\pm$0.011	&	AST3-3	\\		
60095.608	&	12.820	&	$g$	&	11.144$\pm$0.011	&	AST3-3	\\		
60095.625	&	12.837	&	$g$	&	11.130$\pm$0.011	&	AST3-3	\\		
60095.652	&	12.864	&	$g$	&	11.134$\pm$0.011	&	AST3-3	\\		
60095.680	&	12.892	&	$g$	&	11.129$\pm$0.011	&	AST3-3	\\		
60095.691	&	12.903	&	$g$	&	11.159$\pm$0.011	&	AST3-3	\\		
60095.710	&	12.922	&	$g$	&	11.151$\pm$0.011	&	AST3-3	\\	
60096.574	&	13.786	&	$g$	&	11.131$\pm$0.045	&	BOOTES-4	\\		
60096.579	&	13.791	&	$g$	&	11.128$\pm$0.045	&	BOOTES-4	\\		
60096.584	&	13.796	&	$g$	&	11.163$\pm$0.045	&	BOOTES-4	\\
60097.574	&	14.786	&	$g$	&	11.188$\pm$0.011	&	AST3-3	\\		
60097.609	&	14.821	&	$g$	&	11.203$\pm$0.011	&	AST3-3	\\		
60097.626	&	14.838	&	$g$	&	11.199$\pm$0.011	&	AST3-3	\\		
60097.653	&	14.865	&	$g$	&	11.195$\pm$0.011	&	AST3-3	\\		
60098.692	&	15.904	&	$g$	&	11.235$\pm$0.011	&	AST3-3	\\		
60098.711	&	15.923	&	$g$	&	11.249$\pm$0.011	&	AST3-3	\\	
60099.549	&	16.761	&	$g$	&	11.229$\pm$0.027	&	BOOTES-4	\\		
60099.554	&	16.766	&	$g$	&	11.238$\pm$0.027	&	BOOTES-4	\\		
60099.559	&	16.771	&	$g$	&	11.266$\pm$0.027	&	BOOTES-4	\\		
60099.565	&	16.777	&	$g$	&	11.275$\pm$0.027	&	BOOTES-4	\\		
60099.571	&	16.783	&	$g$	&	11.246$\pm$0.027	&	BOOTES-4	\\
60082.847	&	0.059	&	$B$	&	$>18.5$ 		&	10.7-cm	refractor	\\	
60082.876	&	0.088	&	$B$	&	$>18.5$ 		&	10.7-cm	refractor	\\	
60082.889	&	0.101	&	$B$	&	19.006$\pm$0.443	&	12-inch	RC	telescope	\\
60082.906	&	0.118	&	$B$	&	18.632$\pm$0.242	&	12-inch	RC	telescope	\\
60082.944	&	0.156	&	$B$	&	18.228$\pm$0.231 		&	12-inch	RC	telescope	\\
60082.945	&	0.157	&	$B$	&	17.924$\pm$0.190 	&	10.1-cm	refractor	\\	
60082.964	&	0.176	&	$B$	&	17.594$\pm$0.211 		&	15.2-cm	refractor	\\	
60083.024	&	0.236	&	$B$	&	17.826$\pm$0.210 		&	12-inch	RC	telescope	\\
60084.312	&	1.524	&	$B$	&	13.391$\pm$0.152	&	15-cm refractor	\\		
60084.537 & 1.749 & $B$ & 12.83$\pm$0.01 & YAHPT \\
60084.539 & 1.751 & $B$ & 12.81$\pm$0.01 & YAHPT \\
60084.540 & 1.752 & $B$ & 12.83$\pm$0.01 & YAHPT \\
60084.575 & 1.787 & $B$ & 12.76$\pm$0.01 & YAHPT \\
60084.598 & 1.810 & $B$ & 12.71$\pm$0.01 & YAHPT \\
60084.650 & 1.862 & $B$ & 12.62$\pm$0.01 & YAHPT \\
60088.744 & 5.956 & $B$ & 11.11$\pm$0.01 & YAHPT \\
60089.541 & 6.753 & $B$ & 11.20$\pm$0.01 & YAHPT \\
60091.725 & 8.937 & $B$ & 11.28$\pm$0.01 & YAHPT \\
60091.820 & 9.032 & $B$ & 11.31$\pm$0.01 & YAHPT \\
60092.747 & 9.959 & $B$ & 11.34$\pm$0.01 & YAHPT \\
60094.647 & 11.859 & $B$ & 11.40$\pm$0.01 & YAHPT \\
60094.741 & 11.953 & $B$ & 11.39$\pm$0.01 & YAHPT \\
60095.699 & 12.911 & $B$ & 11.41$\pm$0.01 & YAHPT \\
60096.743 & 13.955 & $B$ & 11.45$\pm$0.01 & YAHPT \\
60097.645 & 14.857 & $B$ & 11.51$\pm$0.01 & YAHPT \\
60099.546 & 16.758 & $B$ & 11.59$\pm$0.01 & YAHPT \\
60099.640 & 16.852 & $B$ & 11.60$\pm$0.01 & YAHPT \\
60099.803 & 17.015 & $B$ & 11.58$\pm$0.01 & YAHPT \\
60100.581 & 17.793 & $B$ & 11.65$\pm$0.01 & YAHPT \\

\hline
$^a$Phase relative to first-light time. \\ 
    
\end{longtable}

\begin{table}[!h]
    \centering
    \begin{threeparttable}
    \caption{The best-fit parameters from joint fitting to the $grV$-band early light curves within $\sim$3.0 days after the first detection. The joint fitting assumes the same $t_0$ in the fitting.}
    \label{tab:early_lc}
    \begin{tabular}{ccccccc}
        \hline
        Broken Power-law\\
        \hline
        Band & Data Range\tnote{a} (days) & $t_{0}$(MJD) & $t_{b}$\tnote{b} (days) & $\alpha_1$\tnote{c}  &$\alpha_2$ & Reduced $\chi^2$\\
        \hline
        $g$ & 0.0-2.5 &$60082.63^{+0.07}_{-0.05}$ & $1.18^{+0.15}_{-0.31}$ & $2.40^{+0.31}_{-0.44}$ & $3.45^{+0.11}_{-0.11}$ & 4.3 \\
        $V$ & 0.0-2.5 &$60082.63^{+0.07}_{-0.05}$ & $1.40^{+0.11}_{-0.15}$ & $2.11^{+0.19}_{-0.29}$ & $3.38^{+0.19}_{-0.24}$ & 1.2 \\
        $r$ & 0.0-2.5 &$60082.63^{+0.07}_{-0.05}$ & $1.52^{+0.19}_{-0.14}$ & $2.10^{+0.29}_{-0.30}$ & $3.68^{+0.10}_{-0.15}$ & 4.6 \\
        \hline
        & & & & & & Total 3.4\\
        \hline
        Single Power-law\\
        \hline
        $g$ & 0.4-3.0 & $60082.92^{+0.01}_{-0.01}$ & - & $2.18^{+0.02}_{-0.02}$ & - & 10.1\\
        $V$ & 0.4-3.0 & $60082.92^{+0.01}_{-0.01}$ & - &$2.19^{+0.02}_{-0.02}$ & - & 4.2\\
         $r$ & 0.4-3.0 & $60082.92^{+0.01}_{-0.01}$ & - & $2.25^{+0.02}_{-0.02}$ & - & 12.6\\
        \hline
        & & & & & & Total 10.2 \\
        \hline
    \end{tabular} 
    \begin{tablenotes}
        \footnotesize
        \item[a] The time intervals correspond to the first detection in $r$-band. 
        \item[b] The days since the time of first light $t_0$.
        \item[c] Similar to the exponent index in the single power-law function.
    \end{tablenotes}
    \end{threeparttable}
\end{table}

\begin{table}[!h]
\caption{Best-fit Parameters of the Ejecta$-$CSM Interaction Component.}
    \centering
    
    \begin{tabular}{cc}
    \hline
    Parameters             & Results        \\
    \hline
    $R_{0}$ ($10^{14}$ cm) & $2.0$  \\
    $R_{1}$                & $4.0$ \\
    $R_{2}$                & $6.0$ \\
    $R_{3}$                & $16.0$ \\
    $\dot{M}_{\rm w}(0)$ (M$_{\odot}$ yr$^{-1}$) & $5.0\times10^{-2}$   \\
    $n_{1}$                & $1.5$ \\
    $n_{2}$                & $3.0$ \\
    \hline
    \end{tabular} \\
    
    To match the early light curves of SN\,2023ixf, the radial distribution of CSM ranges from $2.0\times10^{14}$ cm ($R_0$) to $1.6\times10^{15}$ cm ($R_3$).
    \label{tab:interaction}
\end{table}


\clearpage 
\begin{thebibliography}{10}

\bibitem{2018Natur.554..497B}
M.~C. {Bersten}, G.~{Folatelli}, F.~{Garc{\'\i}a}, S.~D. {van Dyk}, O.~G. {Benvenuto}, M.~{Orellana}, V.~{Buso}, J.~L. {S{\'a}nchez}, M.~{Tanaka}, K.~{Maeda}, A.~V. {Filippenko}, W.~{Zheng}, T.~G. {Brink}, S.~B. {Cenko}, T.~{de Jaeger}, S.~{Kumar}, T.~J. {Moriya}, K.~{Nomoto}, D.~A. {Perley}, I.~{Shivvers}, and N.~{Smith}.
\newblock {A surge of light at the birth of a supernova}.
\newblock {\em \nat}, 554(7693):497--499, February 2018.

\bibitem{2008Natur.453..469S}
A.~M. {Soderberg}, E.~{Berger}, K.~L. {Page}, P.~{Schady}, J.~{Parrent}, D.~{Pooley}, X.~Y. {Wang}, E.~O. {Ofek}, A.~{Cucchiara}, A.~{Rau}, E.~{Waxman}, J.~D. {Simon}, D.~C.~J. {Bock}, P.~A. {Milne}, M.~J. {Page}, J.~C. {Barentine}, S.~D. {Barthelmy}, A.~P. {Beardmore}, M.~F. {Bietenholz}, P.~{Brown}, A.~{Burrows}, D.~N. {Burrows}, G.~{Byrngelson}, S.~B. {Cenko}, P.~{Chandra}, J.~R. {Cummings}, D.~B. {Fox}, A.~{Gal-Yam}, N.~{Gehrels}, S.~{Immler}, M.~{Kasliwal}, A.~K.~H. {Kong}, H.~A. {Krimm}, S.~R. {Kulkarni}, T.~J. {Maccarone}, P.~{M{\'e}sz{\'a}ros}, E.~{Nakar}, P.~T. {O'Brien}, R.~A. {Overzier}, M.~{de Pasquale}, J.~{Racusin}, N.~{Rea}, and D.~G. {York}.
\newblock {An extremely luminous X-ray outburst at the birth of a supernova}.
\newblock {\em \nat}, 453(7194):469--474, May 2008.

\bibitem{2008Sci...321..223S}
Kevin {Schawinski}, Stephen {Justham}, Christian {Wolf}, Philipp {Podsiadlowski}, Mark {Sullivan}, Katrien~C. {Steenbrugge}, Tony {Bell}, Hermann-Josef {R{\"o}ser}, Emma~S. {Walker}, Pierre {Astier}, Dave {Balam}, Christophe {Balland}, Ray {Carlberg}, Alex {Conley}, Dominique {Fouchez}, Julien {Guy}, Delphine {Hardin}, Isobel {Hook}, D.~Andrew {Howell}, Reynald {Pain}, Kathy {Perrett}, Chris {Pritchet}, Nicolas {Regnault}, and Sukyoung~K. {Yi}.
\newblock {Supernova Shock Breakout from a Red Supergiant}.
\newblock {\em Science}, 321(5886):223--226, July 2008.

\bibitem{2014Natur.509..471G}
Avishay {Gal-Yam}, I.~{Arcavi}, E.~O. {Ofek}, S.~{Ben-Ami}, S.~B. {Cenko}, M.~M. {Kasliwal}, Y.~{Cao}, O.~{Yaron}, D.~{Tal}, J.~M. {Silverman}, A.~{Horesh}, A.~{De Cia}, F.~{Taddia}, J.~{Sollerman}, D.~{Perley}, P.~M. {Vreeswijk}, S.~R. {Kulkarni}, P.~E. {Nugent}, A.~V. {Filippenko}, and J.~C. {Wheeler}.
\newblock {A Wolf-Rayet-like progenitor of SN 2013cu from spectral observations of a stellar wind}.
\newblock {\em \nat}, 509(7501):471--474, May 2014.

\bibitem{2016ApJ...820...23G}
P.~M. {Garnavich}, B.~E. {Tucker}, A.~{Rest}, E.~J. {Shaya}, R.~P. {Olling}, D.~{Kasen}, and A.~{Villar}.
\newblock {Shock Breakout and Early Light Curves of Type II-P Supernovae Observed with Kepler}.
\newblock {\em \apj}, 820(1):23, March 2016.

\bibitem{2017NatPh..13..510Y}
O.~{Yaron}, D.~A. {Perley}, A.~{Gal-Yam}, J.~H. {Groh}, A.~{Horesh}, E.~O. {Ofek}, S.~R. {Kulkarni}, J.~{Sollerman}, C.~{Fransson}, A.~{Rubin}, P.~{Szabo}, N.~{Sapir}, F.~{Taddia}, S.~B. {Cenko}, S.~{Valenti}, I.~{Arcavi}, D.~A. {Howell}, M.~M. {Kasliwal}, P.~M. {Vreeswijk}, D.~{Khazov}, O.~D. {Fox}, Y.~{Cao}, O.~{Gnat}, P.~L. {Kelly}, P.~E. {Nugent}, A.~V. {Filippenko}, R.~R. {Laher}, P.~R. {Wozniak}, W.~H. {Lee}, U.~D. {Rebbapragada}, K.~{Maguire}, M.~{Sullivan}, and M.~T. {Soumagnac}.
\newblock {Confined dense circumstellar material surrounding a regular type II supernova}.
\newblock {\em Nature Physics}, 13(5):510--517, February 2017.

\bibitem{2022Natur.611..256C}
Wenlei {Chen}, Patrick~L. {Kelly}, Masamune {Oguri}, Thomas~J. {Broadhurst}, Jose~M. {Diego}, Najmeh {Emami}, Alexei~V. {Filippenko}, Tommaso~L. {Treu}, and Adi {Zitrin}.
\newblock {Shock cooling of a red-supergiant supernova at redshift 3 in lensed images}.
\newblock {\em \nat}, 611(7935):256--259, November 2022.

\bibitem{2022ApJ...934L...7R}
Adam~G. {Riess}, Wenlong {Yuan}, Lucas~M. {Macri}, Dan {Scolnic}, Dillon {Brout}, Stefano {Casertano}, David~O. {Jones}, Yukei {Murakami}, Gagandeep~S. {Anand}, Louise {Breuval}, Thomas~G. {Brink}, Alexei~V. {Filippenko}, Samantha {Hoffmann}, Saurabh~W. {Jha}, W.~{D'arcy Kenworthy}, John {Mackenty}, Benjamin~E. {Stahl}, and WeiKang {Zheng}.
\newblock {A Comprehensive Measurement of the Local Value of the Hubble Constant with 1 km s$^{-1}$ Mpc$^{-1}$ Uncertainty from the Hubble Space Telescope and the SH0ES Team}.
\newblock {\em \apjl}, 934(1):L7, July 2022.

\bibitem{2010ApJ...725..904N}
Ehud {Nakar} and Re'em {Sari}.
\newblock {Early Supernovae Light Curves Following the Shock Breakout}.
\newblock {\em \apj}, 725(1):904--921, December 2010.

\bibitem{2011ApJ...728...63R}
Itay {Rabinak} and Eli {Waxman}.
\newblock {The Early UV/Optical Emission from Core-collapse Supernovae}.
\newblock {\em \apj}, 728(1):63, February 2011.

\bibitem{2023MNRAS.522.2764M}
Jonathan {Morag}, Nir {Sapir}, and Eli {Waxman}.
\newblock {Shock cooling emission from explosions of red supergiants - I. A numerically calibrated analytic model}.
\newblock {\em \mnras}, 522(2):2764--2776, June 2023.

\bibitem{2023TNSTR1158....1I}
K.~{Itagaki}.
\newblock {Transient Discovery Report for 2023-05-19}.
\newblock {\em Transient Name Server Discovery Report}, 2023-1158:1, May 2023.

\bibitem{2023TNSAN.119....1P}
D.~A. {Perley}, A.~{Gal-Yam}, I.~{Irani}, and E.~{Zimmerman}.
\newblock {LT Classification of SN 2023ixf as a Type II Supernova in M101}.
\newblock {\em Transient Name Server AstroNote}, 119:1, May 2023.

\bibitem{2023TNSAN.130....1M}
Y.~{Mao}, M.~{Zhang}, G.~{Cai}, J.~{Chen}, J.~{Chen}, X.~{Gao}, K.~{Li}, X.~{Lyu}, Y.~{Qin}, G.~{Sun}, S.~{Xu}, Z.~{Zhang}, J.~{Zhang}, J.~{Zhao}, X.~{Zheng}, W.~{Zhou}, and Q.~{Ye}.
\newblock {Onset of SN 2023ixf observed over East Asian longitudes}.
\newblock {\em Transient Name Server AstroNote}, 130:1, May 2023.

\bibitem{2014ApJ...788..193N}
Ehud {Nakar} and Anthony~L. {Piro}.
\newblock {Supernovae with Two Peaks in the Optical Light Curve and the Signature of Progenitors with Low-mass Extended Envelopes}.
\newblock {\em \apj}, 788(2):193, June 2014.

\bibitem{2017ApJ...838..130S}
Nir {Sapir} and Eli {Waxman}.
\newblock {UV/Optical Emission from the Expanding Envelopes of Type II Supernovae}.
\newblock {\em \apj}, 838(2):130, April 2017.

\bibitem{2023ApJ...954L..42J}
W.~V. {Jacobson-Gal{\'a}n}, L.~{Dessart}, R.~{Margutti}, R.~{Chornock}, R.~J. {Foley}, C.~D. {Kilpatrick}, D.~O. {Jones}, K.~{Taggart}, C.~R. {Angus}, S.~{Bhattacharjee}, L.~A. {Braff}, D.~{Brethauer}, A.~J. {Burgasser}, F.~{Cao}, C.~M. {Carlile}, K.~C. {Chambers}, D.~A. {Coulter}, E.~{Dominguez-Ruiz}, C.~B. {Dickinson}, T.~{de Boer}, A.~{Gagliano}, C.~{Gall}, H.~{Gao}, E.~L. {Gates}, S.~{Gomez}, M.~{Guolo}, M.~R.~J. {Halford}, J.~{Hjorth}, M.~E. {Huber}, M.~N. {Johnson}, P.~R. {Karpoor}, T.~{Laskar}, N.~{LeBaron}, Z.~{Li}, Y.~{Lin}, S.~D. {Loch}, P.~D. {Lynam}, E.~A. {Magnier}, P.~{Maloney}, D.~J. {Matthews}, M.~{McDonald}, H.~Y. {Miao}, D.~{Milisavljevic}, Y.~C. {Pan}, S.~{Pradyumna}, C.~L. {Ransome}, J.~M. {Rees}, A.~{Rest}, C.~{Rojas-Bravo}, N.~R. {Sandford}, L.~Sandoval {Ascencio}, S.~{Sanjaripour}, A.~{Savino}, H.~{Sears}, N.~{Sharei}, S.~J. {Smartt}, E.~R. {Softich}, C.~A. {Theissen}, S.~{Tinyanont}, H.~{Tohfa}, V.~A. {Villar}, Q.~{Wang}, R.~J. {Wainscoat}, A.~L. {Westerling}, E.~{Wiston}, M.~A.
  {Wozniak}, S.~K. {Yadavalli}, and Y.~{Zenati}.
\newblock {SN 2023ixf in Messier 101: Photo-ionization of Dense, Close-in Circumstellar Material in a Nearby Type II Supernova}.
\newblock {\em \apjl}, 954(2):L42, September 2023.

\bibitem{zhang2023circumstellar}
Jujia Zhang, Han Lin, Xiaofeng Wang, Zeyi Zhao, Liping Li, Jialian Liu, Shengyu Yan, Danfeng Xiang, Huijuan Wang, and Jinming Bai.
\newblock Circumstellar material ejected violently by a massive star immediately before its death.
\newblock {\em Science Bulletin}, 68:2548--2554, 2023.

\bibitem{2023ApJ...956...46S}
Nathan {Smith}, Jeniveve {Pearson}, David~J. {Sand}, Ilya {Ilyin}, K.~Azalee {Bostroem}, Griffin {Hosseinzadeh}, and Manisha {Shrestha}.
\newblock {High-resolution Spectroscopy of SN 2023ixf's First Week: Engulfing the Asymmetric Circumstellar Material}.
\newblock {\em \apj}, 956(1):46, October 2023.

\bibitem{2023ApJ...956L...5B}
K.~Azalee {Bostroem}, Jeniveve {Pearson}, Manisha {Shrestha}, David~J. {Sand}, Stefano {Valenti}, Saurabh~W. {Jha}, Jennifer~E. {Andrews}, Nathan {Smith}, Giacomo {Terreran}, Elizabeth {Green}, Yize {Dong}, Michael {Lundquist}, Joshua {Haislip}, Emily~T. {Hoang}, Griffin {Hosseinzadeh}, Daryl {Janzen}, Jacob~E. {Jencson}, Vladimir {Kouprianov}, Emmy {Paraskeva}, Nicolas~E. {Meza Retamal}, Daniel~E. {Reichart}, Iair {Arcavi}, Alceste~Z. {Bonanos}, Michael~W. {Coughlin}, Ross {Dobson}, Joseph {Farah}, Llu{\'\i}s {Galbany}, Claudia {Guti{\'e}rrez}, Suzanne {Hawley}, Leslie {Hebb}, Daichi {Hiramatsu}, D.~Andrew {Howell}, Takashi {Iijima}, Ilya {Ilyin}, Kiran {Jhass}, Curtis {McCully}, Sean {Moran}, Brett~M. {Morris}, Alessandra~C. {Mura}, Tom{\'a}s~E. {M{\"u}ller-Bravo}, James {Munday}, Megan {Newsome}, Maria~Th. {Pabst}, Paolo {Ochner}, Estefania~Padilla {Gonzalez}, Andrea {Pastorello}, Craig {Pellegrino}, Lara {Piscarreta}, Aravind~P. {Ravi}, Andrea {Reguitti}, Laura {Salo}, J{\'o}zsef {Vink{\'o}}, Kellie {de
  Vos}, J.~C. {Wheeler}, G.~Grant {Williams}, and Samuel {Wyatt}.
\newblock {Early Spectroscopy and Dense Circumstellar Medium Interaction in SN 2023ixf}.
\newblock {\em \apjl}, 956(1):L5, October 2023.

\bibitem{2023ApJ...954L..12T}
Rishabh~Singh {Teja}, Avinash {Singh}, Judhajeet {Basu}, G.~C. {Anupama}, D.~K. {Sahu}, Anirban {Dutta}, Vishwajeet {Swain}, Tatsuya {Nakaoka}, Utkarsh {Pathak}, Varun {Bhalerao}, Sudhanshu {Barway}, Harsh {Kumar}, Nayana {A.~J.}, Ryo {Imazawa}, Brajesh {Kumar}, and Koji~S. {Kawabata}.
\newblock {Far-ultraviolet to Near-infrared Observations of SN 2023ixf: A High-energy Explosion Engulfed in Complex Circumstellar Material}.
\newblock {\em \apjl}, 954(1):L12, September 2023.

\bibitem{2023ApJ...955L...8H}
Daichi {Hiramatsu}, Daichi {Tsuna}, Edo {Berger}, Koichi {Itagaki}, Jared~A. {Goldberg}, Sebastian {Gomez}, {Kishalay De}, Griffin {Hosseinzadeh}, K.~Azalee {Bostroem}, Peter~J. {Brown}, Iair {Arcavi}, Allyson {Bieryla}, Peter~K. {Blanchard}, Gilbert~A. {Esquerdo}, Joseph {Farah}, D.~Andrew {Howell}, Tatsuya {Matsumoto}, Curtis {McCully}, Megan {Newsome}, Estefania~Padilla {Gonzalez}, Craig {Pellegrino}, Jaehyon {Rhee}, Giacomo {Terreran}, J{\'o}zsef {Vink{\'o}}, and J.~Craig {Wheeler}.
\newblock {From Discovery to the First Month of the Type II Supernova 2023ixf: High and Variable Mass Loss in the Final Year before Explosion}.
\newblock {\em \apjl}, 955(1):L8, September 2023.

\bibitem{2023ApJ...951L..31B}
Edo {Berger}, Garrett~K. {Keating}, Raffaella {Margutti}, Keiichi {Maeda}, Kate~D. {Alexander}, Yvette {Cendes}, Tarraneh {Eftekhari}, Mark {Gurwell}, Daichi {Hiramatsu}, Anna Y.~Q. {Ho}, Tanmoy {Laskar}, Ramprasad {Rao}, and Peter K.~G. {Williams}.
\newblock {Millimeter Observations of the Type II SN 2023ixf: Constraints on the Proximate Circumstellar Medium}.
\newblock {\em \apjl}, 951(2):L31, July 2023.

\bibitem{2000ApJ...537..796W}
E.~{Waxman} and B.~T. {Draine}.
\newblock {Dust Sublimation by Gamma-ray Bursts and Its Implications}.
\newblock {\em \apj}, 537(2):796--802, July 2000.

\bibitem{2014MNRAS.440.1810M}
Adam~N. {Morgan}, Daniel~A. {Perley}, S.~Bradley {Cenko}, Joshua~S. {Bloom}, Antonino {Cucchiara}, Joseph~W. {Richards}, Alexei~V. {Filippenko}, Joshua~B. {Haislip}, Aaron {LaCluyze}, Alessandra {Corsi}, Andrea {Melandri}, Bethany~E. {Cobb}, Andreja {Gomboc}, Assaf {Horesh}, Berian {James}, Weidong {Li}, Carole~G. {Mundell}, Daniel~E. {Reichart}, and Iain {Steele}.
\newblock {Evidence for dust destruction from the early-time colour change of GRB 120119A}.
\newblock {\em \mnras}, 440(2):1810--1823, May 2014.

\bibitem{2001ApJS..134..263W}
Joseph~C. {Weingartner} and B.~T. {Draine}.
\newblock {Photoelectric Emission from Interstellar Dust: Grain Charging and Gas Heating}.
\newblock {\em \apjs}, 134(2):263--281, June 2001.

\bibitem{2021MNRAS.508.5766I}
Christopher~M. {Irwin}, Itai {Linial}, Ehud {Nakar}, Tsvi {Piran}, and Re'em {Sari}.
\newblock {Bolometric light curves of aspherical shock breakout}.
\newblock {\em \mnras}, 508(4):5766--5785, December 2021.

\bibitem{2022ApJ...933..164G}
Jared~A. {Goldberg}, Yan-Fei {Jiang}, and Lars {Bildsten}.
\newblock {Shock Breakout in Three-dimensional Red Supergiant Envelopes}.
\newblock {\em \apj}, 933(2):164, July 2022.

\bibitem{2023ApJ...957...64S}
Monika~D. {Soraisam}, Tam{\'a}s {Szalai}, Schuyler~D. {Van Dyk}, Jennifer~E. {Andrews}, Sundar {Srinivasan}, Sang-Hyun {Chun}, Thomas {Matheson}, Peter {Scicluna}, and Diego~A. {Vasquez-Torres}.
\newblock {The SN 2023ixf Progenitor in M101. I. Infrared Variability}.
\newblock {\em \apj}, 957(2):64, November 2023.

\bibitem{2023ApJ...953L..16H}
Griffin {Hosseinzadeh}, Joseph {Farah}, Manisha {Shrestha}, David~J. {Sand}, Yize {Dong}, Peter~J. {Brown}, K.~Azalee {Bostroem}, Stefano {Valenti}, Saurabh~W. {Jha}, Jennifer~E. {Andrews}, Iair {Arcavi}, Joshua {Haislip}, Daichi {Hiramatsu}, Emily {Hoang}, D.~Andrew {Howell}, Daryl {Janzen}, Jacob~E. {Jencson}, Vladimir {Kouprianov}, Michael {Lundquist}, Curtis {McCully}, Nicolas~E. {Meza Retamal}, Maryam {Modjaz}, Megan {Newsome}, Estefania {Padilla Gonzalez}, Jeniveve {Pearson}, Craig {Pellegrino}, Aravind~P. {Ravi}, Daniel~E. {Reichart}, Nathan {Smith}, Giacomo {Terreran}, and J{\'o}zsef {Vink{\'o}}.
\newblock {Shock Cooling and Possible Precursor Emission in the Early Light Curve of the Type II SN 2023ixf}.
\newblock {\em \apjl}, 953(1):L16, August 2023.

\bibitem{2023ApJ...952L..23K}
Charles~D. {Kilpatrick}, Ryan~J. {Foley}, Wynn~V. {Jacobson-Gal{\'a}n}, Anthony~L. {Piro}, Stephen~J. {Smartt}, Maria~R. {Drout}, Alexander {Gagliano}, Christa {Gall}, Jens {Hjorth}, David~O. {Jones}, Kaisey~S. {Mandel}, Raffaella {Margutti}, Enrico {Ramirez-Ruiz}, Conor~L. {Ransome}, V.~Ashley {Villar}, David~A. {Coulter}, Hua {Gao}, David~Jacob {Matthews}, Kirsty {Taggart}, and Yossef {Zenati}.
\newblock {SN 2023ixf in Messier 101: A Variable Red Supergiant as the Progenitor Candidate to a Type II Supernova}.
\newblock {\em \apjl}, 952(1):L23, July 2023.

\bibitem{2023arXiv230901389X}
Danfeng {Xiang}, Jun {Mo}, Lingzhi {Wang}, Xiaofeng {Wang}, Jujia {Zhang}, Han {Lin}, and Lifan {Wang}.
\newblock {The Dusty and Extremely Red Progenitor of the Type II Supernova 2023ixf in Messier 101}.
\newblock {\em arXiv e-prints}, page arXiv:2309.01389, September 2023.

\bibitem{2023TNSAN.150....1C}
Vasilii {Chufarin}, Nikolay {Potapov}, Ivan {Ionov}, Stanislav {Korotkiy}, Sergey {Nazarov}, and Kirill {Sokolovsky}.
\newblock {Further constraints on the eruption time of SN 2023ixf in M101}.
\newblock {\em Transient Name Server AstroNote}, 150:1, May 2023.

\bibitem{2023TNSAN.127....1H}
N.~{Hamann}.
\newblock {Pre-Discovery Images of SN 2023ixf on 18th May 2023 21:19:13 UTC}.
\newblock {\em Transient Name Server AstroNote}, 127:1, May 2023.

\bibitem{2013A&A...558A..33A}
{Astropy Collaboration}, Thomas~P. {Robitaille}, Erik~J. {Tollerud}, Perry {Greenfield}, Michael {Droettboom}, Erik {Bray}, Tom {Aldcroft}, Matt {Davis}, Adam {Ginsburg}, Adrian~M. {Price-Whelan}, Wolfgang~E. {Kerzendorf}, Alexander {Conley}, Neil {Crighton}, Kyle {Barbary}, Demitri {Muna}, Henry {Ferguson}, Fr{\'e}d{\'e}ric {Grollier}, Madhura~M. {Parikh}, Prasanth~H. {Nair}, Hans~M. {Unther}, Christoph {Deil}, Julien {Woillez}, Simon {Conseil}, Roban {Kramer}, James E.~H. {Turner}, Leo {Singer}, Ryan {Fox}, Benjamin~A. {Weaver}, Victor {Zabalza}, Zachary~I. {Edwards}, K.~{Azalee Bostroem}, D.~J. {Burke}, Andrew~R. {Casey}, Steven~M. {Crawford}, Nadia {Dencheva}, Justin {Ely}, Tim {Jenness}, Kathleen {Labrie}, Pey~Lian {Lim}, Francesco {Pierfederici}, Andrew {Pontzen}, Andy {Ptak}, Brian {Refsdal}, Mathieu {Servillat}, and Ole {Streicher}.
\newblock {Astropy: A community Python package for astronomy}.
\newblock {\em \aap}, 558:A33, October 2013.

\bibitem{2018AJ....156..123A}
{Astropy Collaboration}, A.~M. {Price-Whelan}, B.~M. {Sip{\H{o}}cz}, H.~M. {G{\"u}nther}, P.~L. {Lim}, S.~M. {Crawford}, S.~{Conseil}, D.~L. {Shupe}, M.~W. {Craig}, N.~{Dencheva}, A.~{Ginsburg}, J.~T. {VanderPlas}, L.~D. {Bradley}, D.~{P{\'e}rez-Su{\'a}rez}, M.~{de Val-Borro}, T.~L. {Aldcroft}, K.~L. {Cruz}, T.~P. {Robitaille}, E.~J. {Tollerud}, C.~{Ardelean}, T.~{Babej}, Y.~P. {Bach}, M.~{Bachetti}, A.~V. {Bakanov}, S.~P. {Bamford}, G.~{Barentsen}, P.~{Barmby}, A.~{Baumbach}, K.~L. {Berry}, F.~{Biscani}, M.~{Boquien}, K.~A. {Bostroem}, L.~G. {Bouma}, G.~B. {Brammer}, E.~M. {Bray}, H.~{Breytenbach}, H.~{Buddelmeijer}, D.~J. {Burke}, G.~{Calderone}, J.~L. {Cano Rodr{\'\i}guez}, M.~{Cara}, J.~V.~M. {Cardoso}, S.~{Cheedella}, Y.~{Copin}, L.~{Corrales}, D.~{Crichton}, D.~{D'Avella}, C.~{Deil}, {\'E}.~{Depagne}, J.~P. {Dietrich}, A.~{Donath}, M.~{Droettboom}, N.~{Earl}, T.~{Erben}, S.~{Fabbro}, L.~A. {Ferreira}, T.~{Finethy}, R.~T. {Fox}, L.~H. {Garrison}, S.~L.~J. {Gibbons}, D.~A. {Goldstein}, R.~{Gommers}, J.~P.
  {Greco}, P.~{Greenfield}, A.~M. {Groener}, F.~{Grollier}, A.~{Hagen}, P.~{Hirst}, D.~{Homeier}, A.~J. {Horton}, G.~{Hosseinzadeh}, L.~{Hu}, J.~S. {Hunkeler}, {\v{Z}}.~{Ivezi{\'c}}, A.~{Jain}, T.~{Jenness}, G.~{Kanarek}, S.~{Kendrew}, N.~S. {Kern}, W.~E. {Kerzendorf}, A.~{Khvalko}, J.~{King}, D.~{Kirkby}, A.~M. {Kulkarni}, A.~{Kumar}, A.~{Lee}, D.~{Lenz}, S.~P. {Littlefair}, Z.~{Ma}, D.~M. {Macleod}, M.~{Mastropietro}, C.~{McCully}, S.~{Montagnac}, B.~M. {Morris}, M.~{Mueller}, S.~J. {Mumford}, D.~{Muna}, N.~A. {Murphy}, S.~{Nelson}, G.~H. {Nguyen}, J.~P. {Ninan}, M.~{N{\"o}the}, S.~{Ogaz}, S.~{Oh}, J.~K. {Parejko}, N.~{Parley}, S.~{Pascual}, R.~{Patil}, A.~A. {Patil}, A.~L. {Plunkett}, J.~X. {Prochaska}, T.~{Rastogi}, V.~{Reddy Janga}, J.~{Sabater}, P.~{Sakurikar}, M.~{Seifert}, L.~E. {Sherbert}, H.~{Sherwood-Taylor}, A.~Y. {Shih}, J.~{Sick}, M.~T. {Silbiger}, S.~{Singanamalla}, L.~P. {Singer}, P.~H. {Sladen}, K.~A. {Sooley}, S.~{Sornarajah}, O.~{Streicher}, P.~{Teuben}, S.~W. {Thomas}, G.~R. {Tremblay},
  J.~E.~H. {Turner}, V.~{Terr{\'o}n}, M.~H. {van Kerkwijk}, A.~{de la Vega}, L.~L. {Watkins}, B.~A. {Weaver}, J.~B. {Whitmore}, J.~{Woillez}, V.~{Zabalza}, and {Astropy Contributors}.
\newblock {The Astropy Project: Building an Open-science Project and Status of the v2.0 Core Package}.
\newblock {\em \aj}, 156(3):123, September 2018.

\bibitem{2016ApJ...830...27Z}
Barak {Zackay}, Eran~O. {Ofek}, and Avishay {Gal-Yam}.
\newblock {Proper Image Subtraction{\textemdash}Optimal Transient Detection, Photometry, and Hypothesis Testing}.
\newblock {\em \apj}, 830(1):27, October 2016.

\bibitem{AUTOPHOT}
S.~J. {Brennan} and M.~{ Fraser}.
\newblock {The Automated Photometry of Transients pipeline (AUTOPHOT)}.
\newblock {\em \aap}, 667:A62, November 2022.

\bibitem{2015AAS...22533616H}
Arne~A. {Henden}, Stephen {Levine}, Dirk {Terrell}, and Douglas~L. {Welch}.
\newblock {APASS - The Latest Data Release}.
\newblock In {\em American Astronomical Society Meeting Abstracts \#225}, volume 225 of {\em American Astronomical Society Meeting Abstracts}, page 336.16, January 2015.

\bibitem{2014zndo.....11813N}
Matthew {Newville}, Till {Stensitzki}, Daniel~B. {Allen}, and Antonino {Ingargiola}.
\newblock {LMFIT: Non-Linear Least-Square Minimization and Curve-Fitting for Python}.
\newblock Zenodo, September 2014.

\bibitem{2020NatMe..17..261V}
Pauli {Virtanen}, Ralf {Gommers}, Travis~E. {Oliphant}, Matt {Haberland}, Tyler {Reddy}, David {Cournapeau}, Evgeni {Burovski}, Pearu {Peterson}, Warren {Weckesser}, Jonathan {Bright}, St{\'e}fan~J. {van der Walt}, Matthew {Brett}, Joshua {Wilson}, K.~Jarrod {Millman}, Nikolay {Mayorov}, Andrew R.~J. {Nelson}, Eric {Jones}, Robert {Kern}, Eric {Larson}, C.~J. {Carey}, {\.I}lhan {Polat}, Yu~{Feng}, Eric~W. {Moore}, Jake {VanderPlas}, Denis {Laxalde}, Josef {Perktold}, Robert {Cimrman}, Ian {Henriksen}, E.~A. {Quintero}, Charles~R. {Harris}, Anne~M. {Archibald}, Ant{\^o}nio~H. {Ribeiro}, Fabian {Pedregosa}, Paul {van Mulbregt}, and {SciPy 1. 0 Contributors}.
\newblock {SciPy 1.0: fundamental algorithms for scientific computing in Python}.
\newblock {\em Nature Methods}, 17:261--272, February 2020.

\bibitem{AST3_22Univ}
Tianrui {Sun}, Xiaoyan {Li}, Lei {Hu}, Kelai {Meng}, Zijian {Han}, Maokai {Hu}, Zhengyang {Li}, Haikun {Wen}, Fujia {Du}, Shihai {Yang}, Bozhong {Gu}, Xiangyan {Yuan}, Yun {Li}, Huihui {Wang}, Lei {Liu}, Zhenxi {Zhu}, Xuehai {Huang}, Chengming {Lei}, Lifan {Wang}, and Xuefeng {Wu}.
\newblock {Antarctic Survey Telescope 3-3: Overview, System Performance and Preliminary Observations at Yaoan, Yunnan}.
\newblock {\em Universe}, 8(6):303, May 2022.

\bibitem{2008ApJ...675..626W}
Xiaofeng {Wang}, Weidong {Li}, Alexei~V. {Filippenko}, Kevin {Krisciunas}, Nicholas~B. {Suntzeff}, Junzheng {Li}, Tianmeng {Zhang}, Jingsong {Deng}, Ryan~J. {Foley}, Mohan {Ganeshalingam}, Tipei {Li}, YuQing {Lou}, Yulei {Qiu}, Rencheng {Shang}, Jeffrey~M. {Silverman}, Shuangnan {Zhang}, and Youhong {Zhang}.
\newblock {Optical and Near-Infrared Observations of the Highly Reddened, Rapidly Expanding Type Ia Supernova SN 2006X in M100}.
\newblock {\em \apj}, 675(1):626--643, March 2008.

\bibitem{AST3_22FrASS}
Tianrui {Sun}, Lei {Hu}, Songbo {Zhang}, Xiaoyan {Li}, Kelai {Meng}, Xuefeng {Wu}, Lifan {Wang}, and A.~J. {Castro-Tirado}.
\newblock {Pipeline for the Antarctic Survey Telescope 3-3 in Yaoan, Yunnan}.
\newblock {\em Frontiers in Astronomy and Space Sciences}, 9:897100, July 2022.

\bibitem{SFFT_Hu2022}
Lei {Hu}, Lifan {Wang}, Xingzhuo {Chen}, and Jiawen {Yang}.
\newblock {Image Subtraction in Fourier Space}.
\newblock {\em \apj}, 936(2):157, September 2022.

\bibitem{SExtractor}
E.~{Bertin} and S.~{Arnouts}.
\newblock {SExtractor: Software for source extraction.}
\newblock {\em A\&AS}, 117:393--404, June 1996.

\bibitem{2016yCat.2336....0H}
A.~A. {Henden}, M.~{Templeton}, D.~{Terrell}, T.~C. {Smith}, S.~{Levine}, and D.~{Welch}.
\newblock {VizieR Online Data Catalog: AAVSO Photometric All Sky Survey (APASS) DR9 (Henden+, 2016)}.
\newblock {\em VizieR Online Data Catalog}, page II/336, January 2016.

\bibitem{matt_craig_2017_1069648_ccdproc}
Matt Craig, Steve Crawford, Michael Seifert, Thomas Robitaille, Brigitta Sip{\H o}cz, Josh Walawender, Z\`e Vin{\'{\i}}cius, Joe~Philip Ninan, Michael Droettboom, Jiyong Youn, Erik Tollerud, Erik Bray, Nathan Walker, VSN~Reddy Janga, Connor Stotts, Hans~Moritz G{\"u}nther, Evert Rol, Yoonsoo~P. Bach, Larry Bradley, Christoph Deil, Adrian Price-Whelan, Kyle Barbary, Anthony Horton, William Schoenell, Nathan Heidt, Forrest Gasdia, Stefan Nelson, and Ole Streicher.
\newblock astropy/ccdproc: v1.3.0.post1, December 2017.

\bibitem{2010AJ....139.1782L_astrometry.net}
Dustin {Lang}, David~W. {Hogg}, Keir {Mierle}, Michael {Blanton}, and Sam {Roweis}.
\newblock {Astrometry.net: Blind Astrometric Calibration of Arbitrary Astronomical Images}.
\newblock {\em \aj}, 139(5):1782--1800, May 2010.

\bibitem{2016A&A...595A...1G}
{Gaia Collaboration}.
\newblock {The Gaia mission}.
\newblock {\em \aap}, 595:A1, November 2016.

\bibitem{2018A&A...616A...1G}
{Gaia Collaboration}.
\newblock {Gaia Data Release 2. Summary of the contents and survey properties}.
\newblock {\em \aap}, 616:A1, August 2018.

\bibitem{2023TNSAN.129....1K}
M.~R. {Kendurkar} and D.~D. {Balam}.
\newblock {Multi-Band Photometric Follow-up of SN 2023ixf}.
\newblock {\em Transient Name Server AstroNote}, 129:1, May 2023.

\bibitem{2023TNSAN.120....1P}
Daniel~A. { Perley} and Ido {Irani}.
\newblock {ZTF Pre-Discovery Forced Photometry of SN 2023ixf}.
\newblock {\em Transient Name Server AstroNote}, 120:1, May 2023.

\bibitem{riess1999there}
Adam~G Riess, Alexei~V Filippenko, Weidong Li, and Brian~P Schmidt.
\newblock Is there an indication of evolution of type ia supernovae from their rise times?
\newblock {\em The Astronomical Journal}, 118(6):2668, 1999.

\bibitem{2013ApJ...778L..15Z}
WeiKang {Zheng}, Jeffrey~M. {Silverman}, Alexei~V. {Filippenko}, Daniel {Kasen}, Peter~E. {Nugent}, Melissa {Graham}, Xiaofeng {Wang}, Stefano {Valenti}, Fabrizio {Ciabattari}, Patrick~L. {Kelly}, Ori~D. {Fox}, Isaac {Shivvers}, Kelsey~I. {Clubb}, S.~Bradley {Cenko}, Dave {Balam}, D.~Andrew {Howell}, Eric {Hsiao}, Weidong {Li}, G.~Howie {Marion}, David {Sand}, Jozsef {Vinko}, J.~Craig {Wheeler}, and JuJia {Zhang}.
\newblock {The Very Young Type Ia Supernova 2013dy: Discovery, and Strong Carbon Absorption in Early-time Spectra}.
\newblock {\em \apjl}, 778(1):L15, November 2013.

\bibitem{2020MNRAS.498...84Z}
Jujia {Zhang}, Xiaofeng {Wang}, Vink{\'o} {J{\'o}zsef}, Qian {Zhai}, Tianmeng {Zhang}, Alexei~V. {Filippenko}, Thomas~G. {Brink}, WeiKang {Zheng}, {\L}ukasz {Wyrzykowski}, Przemys{\l}aw {Miko{\l}ajczyk}, Fang {Huang}, Liming {Rui}, Jun {Mo}, Hanna {Sai}, Xinhan {Zhang}, Huijuan {Wang}, James~M. {DerKacy}, Eddie {Baron}, K.~{S{\'a}rneczky}, A.~{B{\'o}di}, G.~{Cs{\"o}rnyei}, O.~{Hanyecz}, B.~{Ign{\'a}cz}, Cs~{Kalup}, L.~{Kriskovics}, R.~{K{\"o}nyves-T{\'o}th}, A.~{Ordasi}, A.~{P{\'a}l}, {\'A}.~{S{\'o}dor}, R.~{Szak{\'a}ts}, K.~{Vida}, and G.~{Zsidi}.
\newblock {SN 2018zd: an unusual stellar explosion as part of the diverse Type II Supernova landscape}.
\newblock {\em \mnras}, 498(1):84--100, October 2020.

\bibitem{2017ApJ...837L...2A}
Iair {Arcavi}, Griffin {Hosseinzadeh}, Peter~J. {Brown}, Stephen~J. {Smartt}, Stefano {Valenti}, Leonardo {Tartaglia}, Anthony~L. {Piro}, Jos{\'e}~L. {Sanchez}, Brent {Nicholls}, Berto L.~A.~G. {Monard}, D.~Andrew {Howell}, Curtis {McCully}, David~J. {Sand}, John {Tonry}, Larry {Denneau}, Brian {Stalder}, Ari {Heinze}, Armin {Rest}, Ken~W. {Smith}, and David {Bishop}.
\newblock {Constraints on the Progenitor of SN 2016gkg from Its Shock-cooling Light Curve}.
\newblock {\em \apjl}, 837(1):L2, March 2017.

\bibitem{2011ApJ...737..103S}
Edward~F. {Schlafly} and Douglas~P. {Finkbeiner}.
\newblock {Measuring Reddening with Sloan Digital Sky Survey Stellar Spectra and Recalibrating SFD}.
\newblock {\em \apj}, 737(2):103, August 2011.

\bibitem{2013ApJ...769...67P}
Anthony~L. {Piro} and Ehud {Nakar}.
\newblock {What can we Learn from the Rising Light Curves of Radioactively Powered Supernovae?}
\newblock {\em \apj}, 769(1):67, May 2013.

\bibitem{2010ApJ...708..598P}
Anthony~L. {Piro}, Philip {Chang}, and Nevin~N. {Weinberg}.
\newblock {Shock Breakout from Type Ia Supernova}.
\newblock {\em \apj}, 708(1):598--604, January 2010.

\bibitem{2003LNP...598..171C}
R.~A. {Chevalier} and C.~{Fransson}.
\newblock {Supernova Interaction with a Circumstellar Medium}.
\newblock In K.~{Weiler}, editor, {\em Supernovae and Gamma-Ray Bursters}, volume 598, pages 171--194. 2003.

\bibitem{Hu2023_CSMSNeIa}
Maokai {Hu}, Lifan {Wang}, Xiaofeng {Wang}, and Lingzhi {Wang}.
\newblock {Possible circumstellar interaction origin of the early excess emission in thermonuclear supernovae}.
\newblock {\em \mnras}, 525(1):246--255, October 2023.

\bibitem{1984ApJ...285...89D}
B.~T. {Draine} and H.~M. {Lee}.
\newblock {Optical Properties of Interstellar Graphite and Silicate Grains}.
\newblock {\em \apj}, 285:89, October 1984.

\bibitem{1993ApJ...402..441L}
Ari {Laor} and Bruce~T. {Draine}.
\newblock {Spectroscopic Constraints on the Properties of Dust in Active Galactic Nuclei}.
\newblock {\em \apj}, 402:441, January 1993.

\bibitem{2001ApJ...548..296W}
Joseph~C. {Weingartner} and B.~T. {Draine}.
\newblock {Dust Grain-Size Distributions and Extinction in the Milky Way, Large Magellanic Cloud, and Small Magellanic Cloud}.
\newblock {\em \apj}, 548(1):296--309, February 2001.

\bibitem{2022ApJ...931..110H}
Maokai {Hu}, Lifan {Wang}, and Xiaofeng {Wang}.
\newblock {The Effects of Circumstellar Dust Scattering on the Light Curves and Polarizations of Type Ia Supernovae}.
\newblock {\em \apj}, 931(2):110, June 2022.

\bibitem{1988ApJ...329..814D}
Eli {Dwek}.
\newblock {Will Dust Black Out SN 1987A?}
\newblock {\em \apj}, 329:814, June 1988.

\bibitem{2018ApJ...856..146A}
Niloufar {Afsariardchi} and Christopher~D. {Matzner}.
\newblock {Aspherical Supernovae: Effects on Early Light Curves}.
\newblock {\em \apj}, 856(2):146, April 2018.


\bibitem{2013ApJ...779...60M}
 Christopher D. {Matzner}, Yuri  {Levin} and  Stephen {Ro}.
\newblock {Oblique Shock Breakout in Supernovae and Gamma-Ray Bursts. I. Dynamics and Observational Implications}.
\newblock {\em \apj}, 779(1):60, December 2013.


\bibitem{2023ApJ...955L..37V}
Sergiy~S. {Vasylyev}, Yi~{Yang}, Alexei~V. {Filippenko}, Kishore~C. {Patra}, Thomas~G. {Brink}, Lifan {Wang}, Ryan {Chornock}, Raffaella {Margutti}, Elinor~L. {Gates}, Adam~J. {Burgasser}, Preethi~R. {Karpoor}, Natalie {LeBaron}, Emma {Softich}, Christopher~A. {Theissen}, Eli {Wiston}, and WeiKang {Zheng}.
\newblock {Early Time Spectropolarimetry of the Aspherical Type II Supernova SN 2023ixf}.
\newblock {\em \apjl}, 955(2):L37, October 2023.

\bibitem{2019NatAs...3..766H}
Thiem {Hoang}, Le~Ngoc {Tram}, Hyeseung {Lee}, and Sang-Hyeon {Ahn}.
\newblock {Rotational disruption of dust grains by radiative torques in strong radiation fields}.
\newblock {\em Nature Astronomy}, 3:766--775, May 2019.

\end{thebibliography}
\end{document}